\documentclass[preprint,12pt]{elsarticle}

\usepackage{amssymb,longtable,rotating,multirow,amsmath}

\journal{Astronomy and Computing}

\begin{document}

\begin{frontmatter}



\title{Exo-MerCat: a merged exoplanet catalog with Virtual Observatory connection.}

 \author[label1,label2]{E. Alei}
 \author[label1]{R. Claudi}
 \author[label3]{A. Bignamini}
\author[label3]{M. Molinaro}
 \address[label1]{INAF - Osservatorio Astronomico di Padova, Vicolo dell'Osservatorio 5, 35122 Padova, Italy}
 \address[label2]{Dipartimento di Fisica e Astronomia Galileo Galilei, Universit\'a di Padova, Vicolo dell'Osservatorio 3, 35122 Padova, Italy}
 \address[label3]{INAF - Osservatorio Astronomico di Trieste, via Tiepolo 11, 34143, Trieste, Italy}



\begin{abstract}
The heterogeneity of papers dealing with the discovery and characterization of exoplanets makes every attempt to maintain a uniform exoplanet catalog almost impossible. Four sources currently available online (NASA Exoplanet Archive, Exoplanet Orbit Database, Exoplanet Encyclopaedia, and Open Exoplanet Catalogue) are commonly used by the community, but they can hardly be compared, due to discrepancies in notations and selection criteria.
Exo-MerCat is a Python code that collects and selects the most precise measurement for all interesting planetary and orbital parameters contained in the four databases, accounting for the presence of multiple aliases for the same target. It can download information about the host star as well by the use of Virtual Observatory ConeSearch connections to the major archives such as SIMBAD and those available in VizieR. A Graphical User Interface is provided to filter data based on the user's constraints and generate automatic plots that are commonly used in the exoplanetary community.
   With Exo-MerCat, we retrieved a unique catalog that merges information from the four main databases, standardizing the output and handling notation differences issues. Exo-MerCat can correct as many issues that prevent a direct correspondence between multiple items in the four databases as possible, with the available data. The catalog is available as a VO resource for everyone to use and it is periodically updated, according to the update rates of the source catalogs. 
\end{abstract}



\begin{keyword}
(Stars): planetary systems -- catalogues -- Virtual Observatory Tools



\end{keyword}

\end{frontmatter}


\section{Introduction}
\label{sec:intro}
Not even two years passed since the first Hot Jupiter was discovered by  \citet{mayorandqueloz1995} that people used the information on the
handful of planetary discovered up to then to compare them with Solar System and the three PSR B$1957 +12$ planets \citep{wolszczanandfrail1992} making the first comparative analysis of planetary companions \citep{christiansen2018}. On that occasion,   \citet{mazehetal1997} subdivided the limited sample into two classes of eccentricities: the companions with masses smaller than about 5 Jupiter masses have circular orbits, while more massive companions have eccentric orbits. 
Since 1995, about 4000 new confirmed planets have been discovered, demonstrating that it is a very active field of astronomy in continuous growth. Moreover, in the last few years, the activities in this field moved towards the physical characterization of these new worlds, requiring precise knowledge of the main characteristics of already discovered planetary systems and on their stellar and planetary components. However, the careful collection and organization of these exoplanets' main characteristics are necessary for drawing robust, scientific conclusions taking into account the biases and caveats that have gone into their discovery. All this information could be retrieved by a well documented and online available catalog. 

Up to now, there are four large online catalogs in which, even though with various thresholds on different planetary parameters, most of the available information of discovered planets are collected. These databases (DBs) provide also a rich reference set connected to every single planet allowing the retrieval of the original information and the method used by the single research group to obtain the data.  If multiple parameter sets are available for each planet, some of the catalogs can provide a historical archive of the knowledge of the planet parameters as they evolve with time. The most used online catalogs are 
the {\it Exoplanets Encyclopaedia}\footnote{http://exoplanets.eu/} \citep{Schneider2011},  the {\it NASA Exoplanet Archive}\footnote{ https://exoplanetarchive.ipac.caltech.edu/} \citep{Akeson2013}, the {\it Open Exoplanet Catalogue}\footnote{http://www.openexoplanetcatalog.com/}  \citep{Open} and {\it The Exoplanet Data Explorer}\footnote{http://exoplanets.org/} \citep{Wright2011}. In time these catalogs, mostly the {\it Exoplanets Encyclopaedia}, were used to write several statistical works on the different classes of exoplanets \citep[e.g.: ][]{marcyetal2005, udryandsantos2007, winnandfabrycky2015}.
Each catalog will be discussed in Section\ \ref{sec:stateofart}, but it is worth saying that they are different because each catalog considers different criteria to include a new planet in its collection. These criteria are usually based on the physical properties of the planet or statistical thresholds.  

For example, different catalogs use different mass boundaries or include candidate targets in addition to planets described in peer-review papers. 

A lot of planets have been discovered by the radial velocities method. This method, quite efficient in discovering and very good in confirming transiting candidates, while being able to determine the minimum mass of the planets, is dramatically prone to the activity of the star. As matter of fact, for some of claimed planetary companions an analysis of their NIR radial velocity time series resulted in discharging the planetary hypothesis, confirming instead the activity nature of the signal \citep[e.g. TW Hya, BD +20 1790b,][]{figueiramarmieretal2010, figueirapepeetal2010, carleoetal2018}. The stars used as examples are both very young and the previously claimed planets were hot Jupiters which presence was used to discuss the migration theory in young planetary systems \citep{setiawanetal2008, hernanobispoetal2010}. 

Even though this example is dealing with the interpretation of time series, it introduces a maintenance problem: the removal of the planets from the different DBs depends on the frequency of the catalog update which changes based on the research groups that maintain the catalogs.

Catalogs are useful for identifying and examining the broader population of exoplanets, to find relations among the various observables (see e.g. \citet{UlmerMoll2019}). However, particularly with this latter case, caution must be exercised.  To perform robust population analyses, it is necessary to examine carefully the selection effects and biases in the creation of the catalog. Up to now, only \citet{Bashi2018ACatalogs}, in the knowledge of the authors, analyzed from a statistical point of view the impact of the differences among the catalogs, concluding that although statistical studies are unlikely to be significantly affected by the choice of the DB, it would be desirable to have one consistent catalog accepted by the general exoplanet community as a base for exoplanet statistics and comparison with theoretical predictions.

A few efforts in collecting data from different sources have started, such as the Data \& Analysis Center for Exoplanets
(DACE) database\footnote{https://dace.unige.ch} which also offers links to raw data for most targets included in various catalogs. However, no catalogs able to correctly merge the different datasets while correcting nomenclature and coordinate issues appear to be available to the community.

In this paper, we describe our work in creating Exo-MerCat (Exoplanets Merged Catalog), obtained by the extraction of datasets from the four online catalogs to have a consistent DB of exoplanets, in which alias problems, coordinate and other parameters inconsistencies are checked and fixed. Furthermore, we connect the Exo-MerCat to the most important stellar catalogs, using Virtual Observatory (VO\footnote{See http://www.ivoa.net}) aware tools, to complete the retrieval of host stars parameters. We provided also a simple Graphic User Interface for the selection and the visualization of the results. 

The paper is organized in the following way: in Section\ \ref{sec:stateofart} the four online catalogs characteristics are described and the catalogs are compared in Section\ \ref{sec:catcomp}. All the necessary operations to extract Exo-MerCat, the quality check procedures, the standardization, and the treatment of the critical cases are described in Section\ \ref{sec:exomercat}, while its performances are analyzed in Section\ \ref{sec:performance}. Simple science cases are discussed in Section\ \ref{sec:binaries} and Section\ \ref{sec:bd}. Section\ \ref{sec:vo} describes the catalog update procedure as a workflow and its deployment as a set of VO resources. Section\ \ref{sec:GUI} describes the Graphic User Interface and, finally, in Section\ \ref{sec:conclusions} the conclusion are outlined. 

\section{Current state-of-art}
\label{sec:stateofart}
Since the first discoveries, several online tables were built with the results of the different radial velocity and transit surveys. These catalogs, e.g. California and Carnegie Planet search table \citep{butleretal2006}, Geneva Extrasolar Planet search Programmes\footnote{http://obswww.unige.ch/\~naef/planet/geneva\_planets.html}
and the Extrasolar Planets catalog that is the ancestor of  {\it Exoplanets Encyclopaedia}, were workhorse catalogs in which first-hand data from observers were stored. They have not a general-purpose aim. In 2011, with the creation of {\it Exoplanets Encyclopaedia} by \citet{Schneider2011}, the list of discovered planets became a real catalog with planets discovered not only by radial velocity and transit surveys, but also by astrometry, direct imaging, microlensing, and timing, taking into account also unconfirmed or problematic planets. After that, other groups began to maintain general purpose exoplanet catalogs as well. In this section, we describe the characteristics, the requirements, and criteria that characterize each of the main catalogs that are available online today.

\subsection{Exoplanet Encyclopaedia}

The Exoplanet Encyclopaedia  \citep{Schneider2011} (hereafter EU) stores 98 columns containing planetary, stellar, orbital, and atmospheric parameters with uncertainties for all the planet detections already published or submitted to professional journals or announced by professional astronomers in professional conferences, as well as first-hand updated data on professional websites (including candidates from Kepler and TESS space missions). 
Planets or candidates discovered with a large variety of techniques (transit detection, radial velocities, imaging, microlensing, pulsar timing, astrometry) are included.
Due to the larger pool of references, this catalog contains more data than the other archives: any judgment on the likelihood of data is left to the user. Planets are sorted in four categories (Confirmed, Candidate, Retracted, and Controversial): a planet is considered confirmed if claimed unambiguously in a refereed paper or a professional conference.  Rogue planets and interstellar objects are also included. 

In this database, every detected planet whose mass is lower than 60 Jupiter Masses up to 1 sigma uncertainties is stored.

The Exoplanet Encyclopaedia considers also candidates without any estimate of the mass value but with a known radius: they are included in the candidate planets category. 

Both a scientific and editorial board are present to address the peculiar cases and the most important scientific issues that may concern the data. A group of scientists is involved to translate the webpage into multiple languages.


An overview table of all planets belonging to the archive is accessible through the homepage of the Exoplanet Encyclopaedia website. Also, in this case, the table is easily customizable and can be filtered at will. The output is immediately available to download in different file formats.
Every planet has its page, which contains all the available parameters for both the planetary object and the host star, as well as all the bibliographical entries that involve that target.

The Exoplanet Encyclopaedia provides tools easy to customize for histograms and graphs, as well as correlation diagrams between stellar and planetary characteristics. Multiple polar plots that show the distribution of the exoplanet sample in terms of distance from the Solar System is also accessible via the homepage.

It is also a fully VO aware data resource, its contents being deployed through a TAP service (e.g. TOPCAT \citep{2005ASPC..347...29T}) in the form of an EPN-TAP \citep{EPN} compliant core table.

The website includes also a daily updated bibliography of publications, books, theses, and reports concerning exoplanets; a periodically-updated webpage that lists all known planets on an S-type orbit is also present. The team updates other ancillary webpages devoted to the most important instruments and missions, with links to their documentation files or webpage, and to the upcoming conferences and meetings that could be of interest to the exoplanetary community.

Many other tools are also available, such as an ephemeris predictor, a stability tool, and an atmospheric calculator.

\subsection{Exoplanet Orbit Database}

The Exoplanet Orbit Database \citep{Wright2011, Han2014} (hereafter ORG)  includes 230 columns displaying planetary and stellar information, orbital parameters, transit/secondary eclipses parameters, references to observations and fits, of most planets contained in the peer-reviewed literature (up to June 2018), with uncertainties and limits. Kepler Objects of Interests (KOIs), imaging and microlensing targets are retrieved from the NASA Exoplanet Archive and stored in this archive as well, provided they are not already known false positives. This catalog is no longer regularly updated since June 2018.


This archive contains all planets less massive than 24 $M_{Jup}$. Additional requirements are set for imaged planets, whose planet-star mass ratio (including uncertainties) must be smaller than 0.023 (24 $M_{Jup}$ for solar-mass stars), and whose semi-major axis (or projected separation) is lower than $100\ AU \cdot (Mstar/Msun)$.

The archive aims to provide the highest quality orbital parameters of exoplanets rather than providing a complete presentation of every claimed target. The maintainers require that the period measurement has to be certain to at least 15\%: this, together with its lack of recent updates, justifies the overall lower number of confirmed planets included in the catalog.


 In this database $M$ is often set equal to $M\sin{i}$ when the inclination is not known; if neither $M\sin{i}$ nor $M$ are known, mass is calculated using the mass-radius relation shown in \citep{Han2014}.
 
In case of inconsistent host star names, the maintainers choose constellation names, Bayer designations of Flamsteed numbers if available, or rather give ranked priority to GJ numbers, HD numbers, HD numbers, or HIP numbers. The planet's name is then composed of the combination of the stellar name and planet letter.

When a KOI object is validated, its name is replaced by the official Kepler ID. The old KOI notation is stored in the \texttt{OTHERNAME} column. For most candidates, no coordinates are available, most likely because of strict disclosure policies concerning those targets.


The website also hosts the Exoplanet Data Explorer (EDE), an interactive table with plotting tools for all planets included in the database. It allows custom management of the items in the list, by easily adding more columns or by filtering the rows, or by toggling items to be included in the table (e.g. the KOI sample). It also allows the user to download the table.

Every item in the table is linked to an overview page which summarizes all the available parameters for the given planet, together with the relative references.

A plotting tool is also present, to create scatter plots and histograms. Templates of the most common plots are also present, ready to be used or adjusted according to user preferences.

\subsection{NASA Exoplanet Archive}
NASA Exoplanet Archive \citep{Akeson2013} is a database and a toolset funded by NASA to support astronomers in the exoplanet community.  Users are provided with an interactive table of confirmed planets, containing 50 columns of planetary and stellar parameters with uncertainties and limits.
The catalog includes planets or candidates discovered with a the most important detection techniques (transits, radial velocities, direct imaging,  pulsar timing, microlensing, astrometry).


This archive includes and classifies all objects whose mass or minimum mass is less than 30 Jupiter masses and all those objects that have sufficient follow-up observations and validation, to avoid false positives. Free-floating planets are excluded from the sample. All datasets show orbital/physical properties that appear in peer-reviewed publications. 

 Values for both new exoplanets and updated parameters are weekly updated by monitoring submissions on the most important astronomical journals and \texttt{arXiv.org}\footnote{https://arxiv.org/}. In the case of multiple sets of values available in the literature for a given target, the NExScI (NASA Exoplanet Science Institute) scientists decide which reference to set as the default one, depending on the uncertainties and the completeness of the published data sets. In this archive, therefore, internal consistency in each dataset is preferred, rather than a collection of values for different parameters from various references.

In this dataset, some KOI-like objects may however appear. Those are the ones which were at first published as candidates and then confirmed - and their name changed to a Kepler-NNN notation. When the confirmation of a target happens, this archive does not update the name of the target itself, but the planet is included in the confirmed planets dataset. The updated name is stored in the "alias" column. 
KOI objects and candidate planets are stored in a separate table and are subject to further analysis: their status is then updated and, if necessary, the confirmed catalog is updated.

Overview pages for every planet included in the archive are accessible directly from the general table. Such pages collect planetary properties, stellar parameters, light curves, spectra and radial velocity measurements from both space missions and literature. Different sets of data are available, but only one has been selected by the editorial board as the default one, displayed in the overview table. 

Since data values are sorted by reference, it allows the user to compare stellar and planetary physical and orbital values published by different detection methods.
 The dataset of all confirmed planets can be easily downloaded either by browsing or using the corresponding API (application program interface). The table can be downloaded in multiple formats and both rows and columns can be filtered, selecting only the ones the user is interested in. 
 
 Many different sets of data are available on the website, most importantly the cumulative exoplanet archive, the KOI target list, the Threshold-Crossing Events table, as well as data belonging to the major exoplanetary missions. Other noteworthy tools are the ephemeris retrieval software, the periodogram calculator, the observational planning tool, and the transit light curve fitting tool. It is possible to create plots, histograms, or to download pre-generated ones.

\subsection{Open Exoplanet Catalogue}

The Open Exoplanet Catalogue \citep{Open} (hereafter OEC) is an archive based on small XML files, one for each planetary system. Because of its structure, it can easily display planets orbiting a binary (or multiple) star system, and straightforwardly handle exomoons. Each XML file contains up to 42 parameters describing the planet, the host star and the orbital parameters of each system, in addition to uncertainties and upper limits when available. 

No selection criterion is clearly reported in the available documentation.

The catalog is community-driven and open-source, downloadable from GitHub\footnote{https://github.com/} and editable at will. It aims to collect all announced candidates, but it relies on the contributions provided by the users. Anyone can contribute to the archive, by creating pull requests to the remote GitHub repository. The maintainer periodically checks the validity of all updates and only the updates that are believed to be credible are added. All previous versions of the database are available at any point.


This catalog provides links to images of directly imaged planets or artistic impressions of various targets. 
The database is also accessible on a website, the \textit{Visual Exoplanet Catalogue}\footnote{http://exoplanet.hanno-rein.de/}, and it is used by the iOS \textit{Exoplanet} app\footnote{http://exoplanetapp.com/}. 

On the website, separated tables for planets in the habitable zone and planets in binary systems are also provided. The tables are interactive and easy to filter at will. Overview pages for each planetary systems are also accessible: these provide information about the host stars, the planets, as well as graphs that compare the mass of the planets with the masses of the Solar System planets, and the position of the habitable zone of the system compared to the planetary orbits. 

Many ancillary GitHub repositories are available to the user: these allow the user to download free scripts to make plots, to treat XML files and to access data stored in the catalog in Python. Other formats of the whole database, such as ASCII or comma-separated variables, 
are also available for download.

 \begin{table}[]
     \centering
     \begin{tabular}{@{}p{3.8cm}p{9.5cm}@{}}
        \hline\textbf{ Features } & \textbf{Exoplanet Encyclopaedia (EU)} \\\hline 
         \textbf{{Selection Criteria}} &
$M (M\sin{i})< 60\ M_J+ 1\sigma$
\\
         \textbf{Reference} &Peer-reviewed publications, submitted and announced references \\
         \textbf{Target Status} & Confirmed and candidate planets\\
         \textbf{Decision Making} &Scientific and editorial boards\\
         \textbf{Ancillary tools} & interactive tables, graphic tools, planet overview pages, VO connection, binary systems page, bibliography and conferences pages, ephemeris predictor, stability tool, atmospheric calculator \\
         
          \hline\textbf{Features}  & \textbf{Exoplanet Orbit Database (ORG)} \\\hline
           \textbf{{Selection Criteria}} 
 &$M(M\sin{i}) < 24\ M_J$  \\
         \textbf{Reference} & Peer-reviewed publications \\
         \textbf{Target Status} &  Confirmed and candidate planets\\
         \textbf{Decision Making} & Maintainers\\
         \textbf{Ancillary tools} & interactive tables, graphic tools, planet overview pages \\
         \hline\textbf{Features}  & \textbf{NASA Exoplanet Archive (NASA)} \\\hline 
         \textbf{Selection Criteria} &\small{$M(M\sin{i}) < 30\ M_J$ }  \\
         \textbf{Reference} & Peer-reviewed publications \\
         \textbf{Target Status} &Confirmed planets\\
         \textbf{Decision Making} &NExScI team\\
         \textbf{Ancillary tools} &interactive tables, graphic tools, planet overview pages, mission data tables, API, ephemeris predictor, periodogram calculator, observational planning tool, light curve fitting tool \\
         \hline \textbf{Feeatures}& \textbf{Open Exoplanet Catalog (OEC)}\\\hline
         
          \textbf{Selection Criteria} & - \\
         \textbf{Reference} &By commit on GitHub\\
         \textbf{Target Status} & Confirmed and candidate planets\\
         \textbf{Decision Making} & Maintainers\\
         \textbf{Ancillary tools}  & interactive tables, system overview pages, graphic tools, XML/ASCII/csv versions of the archive, open-source updates\\
     \end{tabular}
     \caption{Summary of all interesting features of the various catalogs.}
     \label{tab:criteria}
 \end{table}

\begin{table*}
  \centering
  \caption{Statistics for all catalogs.  The values marked with an asterisk refer to candidate and/or controversial planets; retracted planets were excluded from the analysis. For the ORG catalog, the values in brackets show the statistics made excluding the theoretical mass values, when the result is different. Update: December 14, 2019.}
 \label{tab:statscats}

\begin{tabular}{ l  p{1.8cm} p{1.8cm}p {1.8cm} p{1.8cm}}
\hline\hline
Query & NASA & ORG & OEC & EU\\\hline

Free Floating? & No & No& Yes &  Yes \\
Candidates? & No & Yes & Yes & Yes \\
Stellar Mass & 89 \% & 92 \% & 97 \% & 95 \%\\ 
Stellar Radius& 90 \% & 97 \% & 97 \% & 95 \%\\ 
Stellar Temperature& 93 \% & 97 \% & 92 \% & 93 \%\\ 
Stellar Metallicity& 77 \% & 90 \% & 81 \% & 90 \%\\ 
Stellar Distance & -  & 44 \% & 54 \% & 47 \%\\ 
 Stellar Age  & 57 \% & -&- & 18 \%\\ 
U Magnitude & -  & -  & 1 \% & - \\ 
B Magnitude & - & -  & 35 \% & - \\ 
V Magnitude & -  & 19 \% & 38 \% & 26 \%\\ 
I Magnitude & -  & - & 10 \% & 42 \%\\ 
J Magnitude & -  & 99 \% & 48 \% & 70 \%\\ 
H Magnitude & -  & 99 \% & 48 \% & 70 \%\\ 
K Magnitude & -  & 99 \% & 48 \% & 70 \%\\ 
 Spectral Type  & -  & -& 34 \% & 24 \%\\

Mass Minimum $(M_{Jup})$ & $6\cdot10^{-5}$ & 0 & $6\cdot10^{-5}$  & $2\cdot10^{-6}$  \\
Mass Maximum $(M_{Jup})$& 30 & 22.62 & 263* & 81.9\\
Msini Minimum $(M_{Jup})$& $9\cdot10^{-4}$  & 0 & $4\cdot10^{-3}$* & $5\cdot10^{-4}$  \\
Msini Maximum $(M_{Jup})$& 55.59 & 22.62 (10) & 27.0 & 63.3 \\
Radius Minimum $(R_{Jup})$& $3\cdot10^{-2}$  &  $2\cdot10^{-2}$* & $2\cdot10^{-3}$*  & $2\cdot10^{-6}$ \\
Radius Maximum $(R_{Jup})$& 6.9& 9730* & 6 & 4332.12*\\\hline

\end{tabular}

\end{table*}

\section{Catalog Comparison}
\label{sec:catcomp}

As reported in \citet{Bashi2018ACatalogs}, the four catalogs are indeed similar, but not equal. 

In Table \ref{tab:criteria} we reported a summary of the features and the selection criteria of each catalog. This is the amount of information we managed to collect by reading the various documentation links provided by the websites. Often, the files lack update, or the main documentation is represented by the release paper itself, which may not consider all of the actual features of the various websites, nor any modification to the DBs themselves. For this reason, we studied carefully the actual boundaries of the various mass and radius parameters for all catalog, shown in Table \ref{tab:statscats}. 

Since some retracted planets appeared in various catalogs (e.g. OEC and EU archives), we excluded them from further analysis. In doing so, it appears clear that all archives follow the preferred selection criterion, when stated in the documentation. Some  extremely high values of planetary radii are present (in ORG and EU archives in particular), belonging to planets labeled as unconfirmed in the various archives.

This discrepancy in the choice of the upper mass boundary in the catalogs is probably linked to the ongoing discussion concerning the mass threshold for which the object is no longer a planet, but a brown dwarf (see Section \ref{sec:bd}).

For what concerns the amount of stellar data present in the various archives, shown in Table \ref{tab:statscats}, we noticed that the overall information about the host star's mass, radius, temperature, and metallicity is fairly complete for all catalogs. All archives but NASA have also magnitudes measurements, even though not all wavelength bands are uniformly filled by the various DBs. Distances, spectral types, and ages information are not provided by all catalogs uniformly. This is, in any case, not so important, since the main goal of such archives is to provide suitable information concerning exoplanets rather than their host stars. Such lack in stellar data can be overcome by looking for more specific and trustworthy data into dedicated catalogs (e.g. SWEET-Cat, \citet{Sweetcat}, other than the most famous stellar catalogs).

A more important analysis can be made on the available planetary measurements for all catalogs. We expect that, because of the different philosophies on the consistency of the datasets, the amount of data available for each target could be different, thus leading to substantially different records for a single planet. 

As shown in Table \ref{tab:stats},  ORG and EU catalogs include a massive amount of candidates, and therefore appear to be much larger than the NASA and OEC archives.
The number of confirmed planets is similar for NASA and EU catalogs, while OEC and ORG archives show fewer items, due either to selection criteria or lack of update. In the OEC and EU catalogs, a handful of planets labeled as false positives are present in the downloaded tables.

In the ORG and EU catalogs, large importance is given to radius and period measurements, while the EU catalog alone seems to be the most complete for what concerns mass and minimum mass. The majority of the mass or minimum mass measurements in the ORG catalog are, as a matter of fact, theoretical.

In all catalogs but the OEC archive, simultaneous values of mass and minimum mass appear for the same target; on the other hand, the majority of the planets having at least one mass-related measurement and a non-null radius value, has a non-null period measurement as well. We expect all transiting targets to fall into this subset. 
By counting all unique host star names in the various archives, we estimated the number of planetary systems as well. This value is not the same for all catalogs, but it reflects the difference in the number of entries in each archive, due to the presence of candidates in some catalogs rather than others.

\begin{table*}
  \centering
  \caption{Available measurements for various combinations of parameters in the four catalogs as they were downloaded from their sources. For the ORG catalog the values in brackets show the statistics made excluding the theoretical mass values, when the result is different.  See  \citet{Bashi2018ACatalogs} for comparison. Update: December 14, 2019. }
 \label{tab:stats}
\begin{tabular}{ l  c c c c }
\hline\hline
Query & NASA & ORG & OEC & EU\\\hline

All planets & 4055 & 5747 & 3846 & 6877 \\
Confirmed & 4055 & 3236 & 3725 & 4149 \\
Candidates & 0 & 2511 & 110 & 2718 \\
False Positives & 0 & 0 & 11 & 10 \\
With radius & 3142 & 4999 & 2959 & 5505 \\
With mass & 877 & 5607 (456)& 1163 & 1314 \\
With msini & 769 & 970 (29)& 281 & 1036 \\
With period & 3941 & 5733 & 3721 & 6603 \\
With mass or msin(i) & 1613 & 5608 (480)& 1444 & 2216 \\
With mass and msin(i) & 33 & 969 (5)& 0 & 134 \\
With mass and msin(i) and radius & 24 & 385 (4) & 0 & 86 \\
With mass or msin(i) and radius & 717 & 4996 (420) & 578 & 931 \\
With mass or msin(i), radius and period& 707 & 4996 (420)& 564 & 902 \\
All systems &  3019 & 4717 & 2852 & 5513 \\\hline

\end{tabular}

\end{table*}

\begin{figure*}
 \centering
 \includegraphics[width=\linewidth]{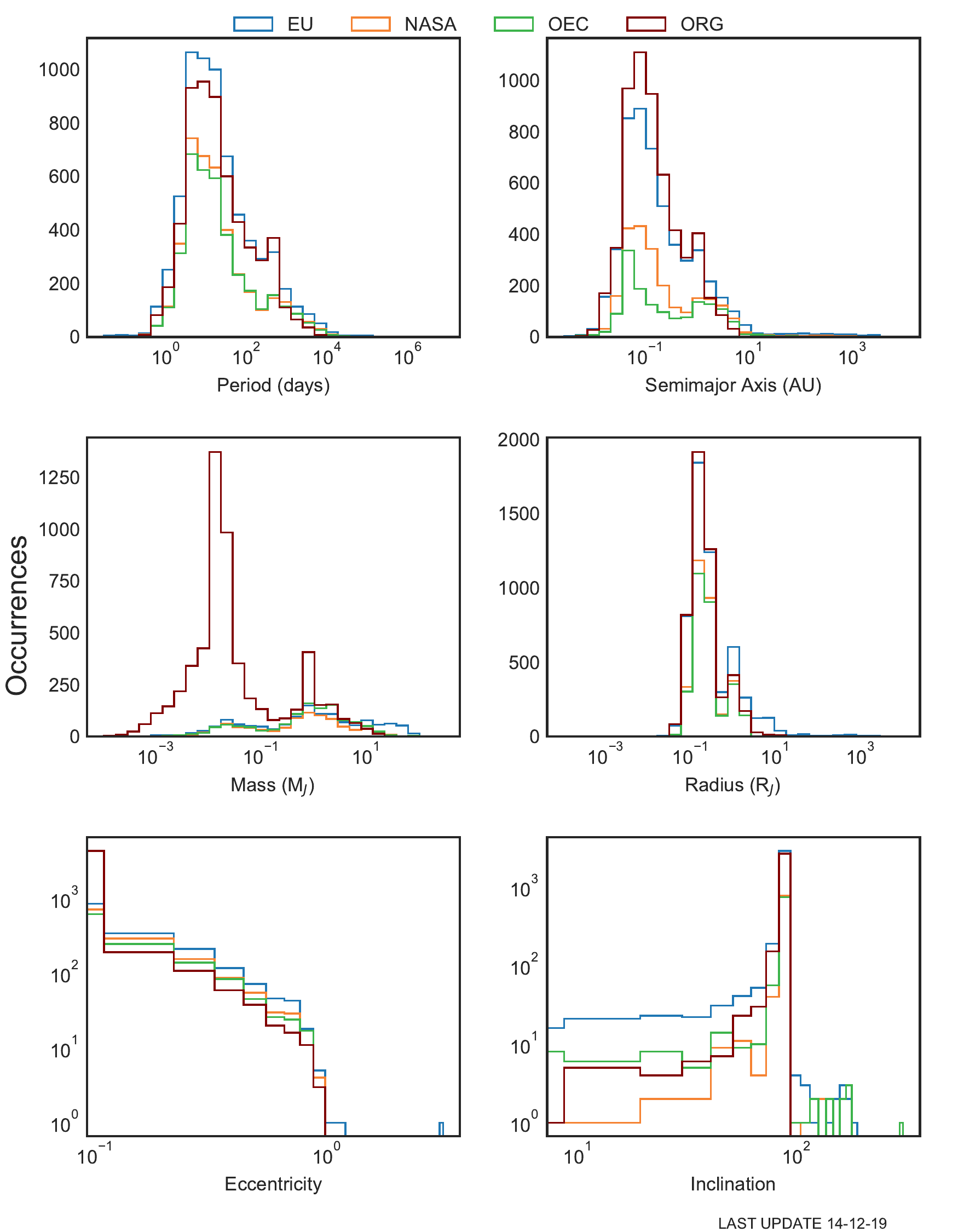}
 \caption{Period, semi-major axis, mass, radius, eccentricity, inclination histograms for each of the input catalogs. As shown in the legend, the blue histograms refer to the Exoplanet Encyclopaedia, the orange histograms to the NASA Exoplanet Archive, the green histograms to the Open Exoplanet Catalogue, and brown ones to the Exoplanet Orbit Database. }
 \label{figure:Multi}
\end{figure*}

We report in Figure \ref{figure:Multi} the distribution of all planetary parameters for the four catalogs. 
Different behavior can be seen from panel to panel. Due to the presence of candidates, the EU and ORG catalogs show higher values of period, semi-major axis, and radius -- which are indeed the first measurable parameters in transiting candidates. The ORG catalog shows also many values of mass, due to the presence of theoretical values. 

No substantial difference is seen in the other graphs. A few uncommon values for the inclination were found in the OEC catalog, probably due to unreliable measurements or theoretical values.

\begin{figure*}
 \centering
 \includegraphics[width=\linewidth]{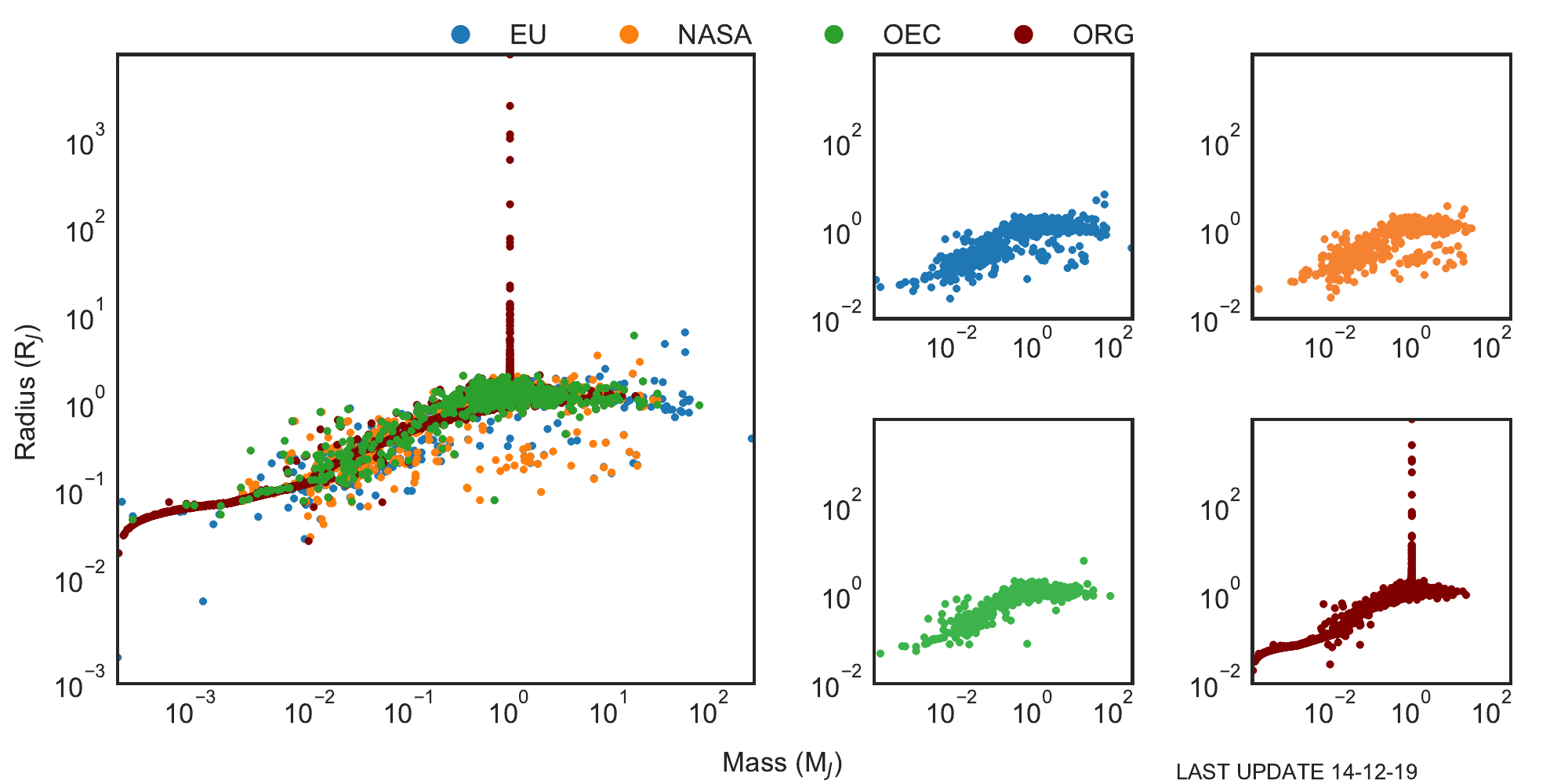}
 \caption{Mass-Radius plot for the raw catalogs. As shown in the legend, blue dots refer to the Exoplanet Encyclopaedia, the orange dots to the NASA Exoplanet Archive, the green dots to the Open Exoplanet Catalogue, and the brown ones to the Exoplanet Orbit Database. }
 \label{figure:MR}
\end{figure*}

The large difference in the number of mass measurements visible in the center-left panel of Figure \ref{figure:Multi} reflects also in the mass-radius plots in Figure \ref{figure:MR}. Even though overall, there seems to be a good agreement among the various measurements, there is indeed a fraction of planets not belonging to all catalogs. While the region around 1-10 Jupiter Masses and 1 Jupiter Radii seems to be more or less equally populated by the four DBs, the area around 1-10 Jupiter masses and 0.1 Jupiter radii is not uniformly covered. On the other hand, the ORG catalog provides a few targets at low masses and Jupiter-like radii, which are absent in the other DBs. Also, a clear trend determined by all mass values retrieved from the theoretical M-R relationship is present in the ORG data. The masses indeed follow the trend determined by observed values, except for the strong vertical at 1 $M_J$ showing that for radii larger than the Jupiter radius, the relation is out of its range of validity.

From these plots, it is clear that any attempt to fully merge the four catalogs is impossible. What we felt the need to do, though, is to provide the four datasets with a greater uniformity, which may lead to a more effective association among the various targets and a higher statistical significance on the measurements, creating a catalog that would cross-match at best the four archives.

\section{Genesis of Exo-MerCat}
\label{sec:exomercat}
Exo-MerCat is a program written in Python 3.6 that merges the exoplanet catalogs described in the previous sections.
To merge the exoplanets catalogs some preliminary operations are necessary, among which the standardization of the four data sets to be able to compare each entry of a catalog with those of the others. This task is very difficult to do automatically and we had to choose in a very accurate way the software tools more suitable for the purposes. 

\begin{figure}
 \centering
 \includegraphics[width=\textwidth]{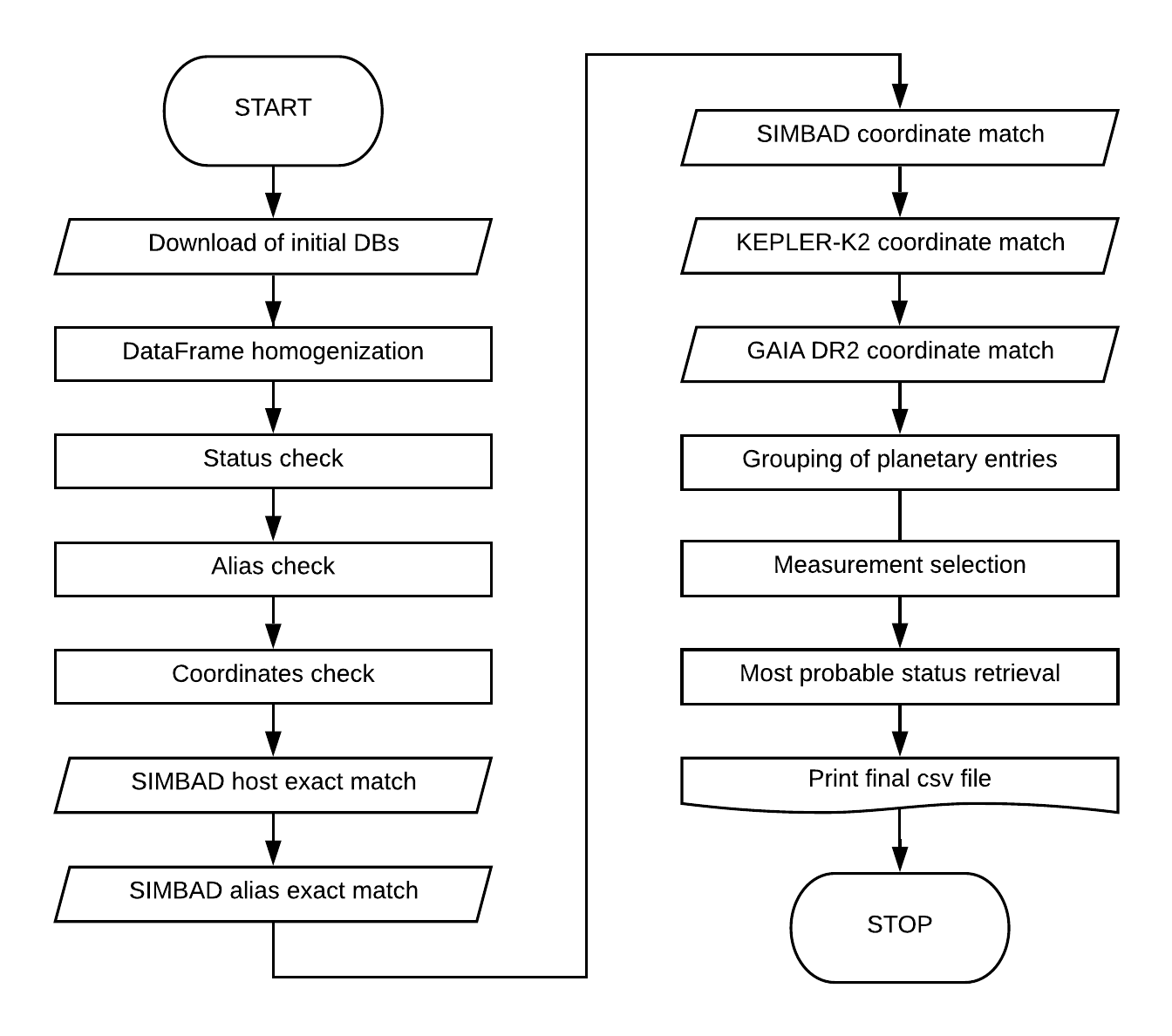}
 \caption{Flowchart of the main script.}
 \label{figure:flow}
\end{figure}

One of the biggest challenges was to hunt for the aliases and check the coordinates of host stars. Most of the aliases problems derived by discrepancies in the notation of both stars and planets in the different catalogs. The flowchart of this software program is reported schematically in Figure\ \ref{figure:flow} and discussed in detail in the following sections.

A Graphical User Interface is provided to all users and it allows the filtering of the catalog, as well as the automatic plotting of some interesting plots.

\subsection{Libraries and Tools}\label{sec:tools}

To be operative, the software needs a few Python packages in addition to the default ones. 
The package \texttt{pandas}\footnote{https://pandas.pydata.org/} allows flexibly manipulating large datasets, by storing data in Series (1-D arrays) or DataFrames (2-D arrays) structures. It also allows data grouping and merging, as well as quick operations between rows and columns, and hierarchical indexing.

The package \texttt{astropy}\footnote{http://www.astropy.org/} is already included in the \texttt{Anaconda} Python Distribution, a community-developed core Python package for Astronomy \citep{astropy:2013, astropy:2018}. In our case, this package was used to treat astronomical coordinates and to properly convert the various parameters.  Also, we used \texttt{astroquery}, an \texttt{astropy} affiliated package, to access and download the original ORG catalog.

For what concerns the Open Exoplanet Catalogue, an \texttt{.xml} reader package is needed. This is by default available in Python, while the retrieval code (which converts an \texttt{.xml} file to a \texttt{pandas} Series) was adapted from the default ones, available at the original website\footnote{https://github.com/hannorein/open\_exoplanet\_catalogue}. 

All the other VO queries were performed using \texttt{pyvo}\footnote{https://pyvo.readthedocs.io/}, an \texttt{astropy} affiliated package, which implements general methods for discovery and access of astronomical data available from archives complying with the standard protocols defined by the International Virtual Observatory Alliance (IVOA).

The software makes extensive use of the Table Access Protocol \citep[TAP,][]{2010ivoa.spec.0327D}, an IVOA standard designed to provide access to relational table sets specifically annotated for astrophysical usage.
The queries posted to TAP compliant services can be specified using the Astronomical Data Query Language \citep[ADQL,][another IVOA standard]{2008ivoa.spec.1030O}. The SQL-like queries built in ADQL and posted to TAP services allow catalog filtering using lists of astronomical targets, as well as spatial cross-matching functions among various catalogs and general custom manipulation of the content of each catalog.

\subsection{Initial datasets retrieval}\label{sec:raw}

There is no uniform retrieval of the raw datasets since not all catalogs allow the same service to download the source file.

For the Exoplanet Orbit Database, the Exoplanet Encyclopaedia, and the NASA Exoplanet Archive, a simple call to command-line instruction \texttt{wget} allows downloading a comma-separated value file. The code selects specific columns when making the \texttt{wget} call, to reduce the amount of downloaded data and to be sure that all necessary columns are correctly considered.

The Open Exoplanet Catalogue is on the other hand composed by a set of separate \texttt{.xml} files, which can be downloaded from the GitHub repository. In this case, the code needs to download the latest updates of the repository itself, and then to convert the \texttt{.xml} files into a unique \texttt{.csv} file.

The various input datasets are stored in four \texttt{pandas DataFrame} objects.

\subsection{Standardization}
\label{sec:standard} 

The raw datasets present themselves as very different, so any sort of merging at this stage would be impossible. For this reason, the DataFrame objects need to be carefully standardized to the desired, common output.

For every single catalog, a dedicated function within the software can process the following operations:
\begin{itemize}
\item First of all, only part of the available columns was considered: at this first step, we chose to focus on the planetary parameters, discarding all information about the host stars since it could be easily retrieved by connecting to the most important stellar catalogs. This choice contributes to the loss of coherence of the final output: it is, however, worth reminding that the philosophy behind Exo-MerCat is to collect as much data as possible from different sources, without any constraint on the homogeneity of references for each dataset. Other catalogs are available to provide information on planet-bearing stars coherently with the planetary reference paper (e.g. SWEET-Cat, \citet{Sweetcat}).
In any case, precise measurements of stellar parameters are not always present in the exoplanets-related references, so columns in the raw databases concerning those are often far from completion.

The parameters taken into account at this stage are (see \ref{app:vodetails} for further information): the values, the errors, and the references of all mass, minimum mass, radius, period, semi-major axis, eccentricity and inclination measurements for every planet in the various catalogs; the planetary and stellar names; the alternative nomenclature strings; the year and method of discovery; any information on the binary nature of the host stars and the status of each planet. When present in the input catalogs, any of the columns storing additional information (stellar mass, age, temperature, radius, distance, magnitudes, transit, and radial velocity parameters) are at present not considered. 

\item Selected columns were then renamed to ease the subsequent merging. For each parameter \texttt{X} (mass, minimum mass, radius, period, semi-major axis, eccentricity, inclination), the code creates a new column called \texttt{X-REF} to store the link to the bibliography in which the measurement first appeared. This was not possible for the "EU" and "OEC" catalog, which do not provide information concerning the reference in the first place; the software keeps track of those by filling the \texttt{X-REF} columns with a string displaying either EU or OEC respectively.

\item A column dedicated to the aliases of a single target is created, displaying them as a comma-separated string. The list of aliases is seldom complete, since most reference papers report up to two aliases per target, despite being it known with other identifiers as well.

\item All double and unnecessary white spaces were removed. 

\item All target names were checked and standardized: all Kepler-like entries were labeled as "Kepler-X" (with X as a 1-4 ciphers integer with no leading zeros), the Greek letters for some stars were displayed as three-character strings ($\alpha$ as alf, $\beta$ as bet...). The host constellations were displayed as three-character strings too. A dictionary of all abbreviations for constellations was retrieved from the IAU official list\footnote{https://www.iau.org/public/themes/constellations/} and use to make coherent replacements.
This step was necessary to allow the merging of stars belonging to a known constellation, but stored in the various databases using a slightly different notation. A suitable example would be the host star Algieba, gamma Leonis. The planet orbiting that star is labeled "gamma 1 Leo b" in the Exoplanet Encyclopaedia, "gamma Leo A b" in the Exoplanet Orbit DataFrame, "gam 1 Leo b" in the NASA Exoplanet Archive, and "Gamma Leonis b" in the Open Exoplanet Catalogue. For a human being, it could be easy to assume that the four entries represent the same target, but a software program that could only compare them as strings would recognize them as undoubtedly different.

\item Generally, a planet is labeled as the name of its host star, plus a letter (b to h) to rank planets within the same stellar system based on the year of discovery. To retrieve the host star name from the catalog, it is necessary to strip the last letter from each target name. On the other hand, unconfirmed Kepler Objects of Interest have a different notation concerning confirmed exoplanets: they are usually displayed as KOI-NNNN.DD where KOI-NNNN represents the host star, while the last two digits DD unambiguously identify each target within the same system (where 01 is the first discovered planet, 02 the second one, etc.). In this case, the last three characters ".DD" were removed from the planet name to retrieve its host star name.

Another exception to handle was represented by the very first exoplanets discovered orbiting the pulsar PSR 1257+12 \citep{wolszczanandfrail1992}. This system was originally labeled as  PSR 1257+12 A,  PSR 1257+12 B, and  PSR 1257+12 C, but since then a massive variation in common notation happened, so we felt the need to change those names to a more standardized  PSR 1257+12 b,  PSR 1257+12 c, and  PSR 1257+12 d so that the planets could be labeled uniformly throughout the final catalog.

\item  For what concerns the labeling of planets orbiting binary systems, the four catalogs behave differently. NASA and EU catalogs provide the letter labeling the host binary companion as a substring in the string displaying the name of the planet; all planets on a P-type orbit (i.e. circumbinary planets) show the substring \texttt{AB} in the name string. The ORG catalog provides a \texttt{BINARY} column that indicates whether the host star is supposed to be part of a binary/multiple system, but it does not provide information concerning the orbit type (whether S- or P-type). This information is on the contrary provided by the OEC catalog within the \texttt{binaryflag} (which is 2 if the planet is on an S-type orbit, 1 if it is on a P-type orbit, 0 otherwise), but little information is given concerning S-type planets since it is not known which stellar companion is the actual host star. 
However, the OEC and ORG catalogs provide as well the substring labeling the binary star (or both of them if circumbinary), but that is often not coherent with the flags. In particular, the ORG catalog provides the letter of the binary companion that hosts a planet only 15 times throughout the whole catalog, but the \texttt{binary} flag indicates that more than 700 planets orbit a binary star (corresponding to nearly 500 unique host star names). On the other hand, the OEC catalog provides about 200 non-null \texttt{binaryflag} values, but less than half of the sample displays the binary substring within the name of the planet. 

The ideal setup for the four DataFrames, to provide a correct match in the following functions, would be to have both information concerning the orbit of the planet, and the binary system that hosts him. This was not possible with the data provided by the catalogs. The software collects as much information as it can from the original datasets by stripping the label letter(s) from the \texttt{host} string, when available and storing this substring in a dedicated column \texttt{binary}, whose value is left empty if no information is provided, assuming that in that case, the host star is a single star. We chose to not take into account the flag value for these two catalogs, due to the incompleteness of the information provided. The only exception was the circumbinary sample in the OEC catalog (\texttt{binaryflag}=2), for which we forced the \texttt{binary} value to be "AB".

\item The letter labeling each planet is stored in a dedicated column that will allow hierarchical indexing of the DataFrame. For Kepler Objects of Interest, the software converts the ciphers DD in letters (01 as b, 02 as c...), to keep a uniform notation among confirmed and unconfirmed objects.
\item In the input catalogs, calculated values for mass and radius can be identified by a flag in dedicated columns. These values could be either retrieved by theoretical mass/radius relations, or by assuming a typical value for the unknown inclination. We chose to set to undefined all values that were calculated or theoretical, thus retaining only actual measured values for these parameters.

\item Finally, the names of the retrieval methods were standardized, since the various catalogs adopted different notations. 
\end{itemize}

\subsection{KOI Objects Status} \label{sec:koi}

 It may be possible that some additional candidates or false positives are included in the current archives, due to lack of updates or human error. A check on the status of each target (especially for Kepler ones, since they represent the majority of known exoplanets) is due.

NASA Exoplanet Archive and Mikulski Archive for Space Telescopes\footnote{https://archive.stsci.edu/index.html} (MAST)  provide an updated table of all Kepler Objects of Interest both belonging to Kepler and K2 missions, periodically updated to show the status of each target, whether confirmed by follow-up observations, still candidate or retracted as false positive. 
For all confirmed planets, a Kepler-like identifier is given to replace the original KOI- or KIC/EPIC-like notation. 

The software downloads the table and cross-matches it with the four DataFrames, updating any KOI name with the official Kepler identifier, if present. Then, it stores the various information concerning the status of each target in a column named \texttt{status}, filling it with \texttt{CONFIRMED}, \texttt{CANDIDATE}, or \texttt{FALSE POSITIVE} strings.

In the best possible scenario, we should not see variations in the number of confirmed, candidates, and false positives as reported in Table \ref{tab:statscats}, thus meaning that the original status of every target in each catalog is correctly updated. This was unfortunately not the case: in the NASA catalog 15 candidates and 1 false positive appeared; in the ORG catalog 1 confirmed planet was actually still a candidate; in the OEC catalog 10 confirmed planets were still candidates or false positives; finally, in the EU catalog nearly 500 false positives were contained in the candidate sample.

This could have been caused by delays in the update of the single catalogs, or either misinterpretation of reference papers. In any case, the NASA Exoplanet Archive appears to be the most updated catalog from this point of view.

In the case where no coordinates for the Kepler candidates are available (i.e. for the ORG catalog), the crossmatch among the exoplanetary DataFrames and the MAST KOI table is useful in retrieving the missing information. The function successfully retrieved all coordinates of the ORG candidates, about 2500 at the time of writing. 

We expect to modify this routine soon, as soon as more TESS candidates will be confirmed by follow-up observations. Provided that a KOI-like table is available for TOI (TESS Objects of Interest) objects, a similar feature will help to treat such targets.

\subsection{Alias and Coordinates Check}\label{sec:checks}

By trying to merge the four DataFrames, we expect to find a large number of targets in two or more catalogs. It may be possible, however, that some targets are labeled differently despite being the very same objects, since the various catalog maintainers may have chosen a different alias to represent the host stars, and thus the orbiting planets. In this way, a code that performs a match among strings would not be effective, not considering all the occurrences of a given planet (see Section \ref{sec:performance}).

For this reason, this function stores all the available host star default names and aliases from the four original DataFrames and attempts to find if (and when) the same host star is saved with an alias. To be more coherent and to ease the way for the operations to come, we would prefer to retain a host star name which can be easily recognized by SIMBAD\footnote{http://simbad.u-strasbg.fr/simbad/} \citep{2000A&AS..143....9W}; furthermore, a more exhaustive list of identifiers for which each stellar target is known would undoubtedly lead to more effective results in this subroutine and the following ones.

All host star names retrieved from the four DataFrames (dropping all duplicated strings)  are therefore queried to SIMBAD using a VO ConeSearch \citep{2008ivoa.specQ0222P} query. For each queried string, the ConeSearch returns a string listing all available aliases, as well as a single string labeling the main identifier for which each target is known in this archive. At the time of writing, starting from a list of about 6300 host star strings (which may contain duplicates of the same physical target labeled differently), only about 550 queries were unsuccessful. This was probably caused by an unconventional notation displayed in this target, mainly the usage of unknown aliases.

SIMBAD can recognize many of the aliases under which a star is known and all of them point to the same target, identified by a unique name. We, therefore, expect that the result of a ConeSearch of the same target queried under different aliases should return the very same results (i.e. main identifier and list of all known aliases). This feature allowed us to further identify duplicates within the list of host star targets. 

When such issues happen, the function chooses a common identifier for each host star and overwrites the host star name in the original DataFrames when necessary. At the time of writing, 320 duplicates within the total list of host star names were found. 

Many of these were KOI-like objects: as a matter of fact, for the Kepler systems in which one or more planets are confirmed (and thus renamed in a Kepler-like notation) while others are still candidates, the host star is by construction named differently. This function, therefore, helps to correct and to uniform such cases, too.  

At this point, the identifier list retrieved for each successful target by SIMBAD was completed if necessary with the available aliases for each star available in the catalogs.

For the host stars for which the VO ConeSearch was unsuccessful, the code performs a less effective yet useful check. For all available identifiers of a target, whether belonging to SIMBAD or saved by the original catalogs, the function queries the Host column of the DataFrame for other occurrences. If an alias in that column is found (i.e. that entry is the same host star but labeled in an alternative way), the host star name is uniformed. At the time of writing, about 20 further corrections were made.

Subsequently, the software checks for the consistency of coordinates, to avoid mismatches when merging the catalogs. Indeed, it may be possible to have coordinate values which are not correctly updated with new measurements, or either sign errors may occur. 

On the other hand, J2000 coordinate differences can be very important in correctly identifying any planet orbiting a binary, especially for those cases in which no label was provided by default. In particular, the same binary companion can appear with the same \texttt{host} name in more than one catalog, but in some cases the \texttt{binary} string would be null (i.e. no information concerning the fact that the host star was part of a binary/multiple system was given by one or more catalogs): in this case, a code which compares strings would interpret the various entries as different targets, even though the actual planet would be the same. Whenever possible, then, this check identifies all targets having different values of the \texttt{binary} string, for each system in the catalog. The software creates subgroups depending on the value of \texttt{binary} (typically "A", "B", "AB", or null) and checks if each pair of coordinates of the null subsample can match any of the coordinates of the other subgroups. In this way, most of the originally null  \texttt{binary} values are fulfilled with the correct value, thus allowing the following functions to perform correct operations among targets. 

Sometimes, the difference in coordinates can be high enough (greater than 0.005 degrees) to forbid an automatic match between the various entries. In this case, the flag  \texttt{MismatchFlagHost} was set as 1 for all the involved targets to warn the user about this issue. 

Furthermore, it may happen that within the same system, S-type and P-type orbiting planets existed simultaneously, depending on the original value of \texttt{binary}, which often belongs to different catalogs. This is a somewhat difficult problem for what concerns the dynamics of the system that needs to be studied carefully. In such cases, it is highly probable that the different entries are in truth the same planet, but two or more catalogs were not in agreement for what concerns the orbital type. We reported such cases by filling the \texttt{MismatchFlagHost} flag as 2.

At the time of writing, about 118  \texttt{binary} corrections were successfully made. Two planetary systems had the  \texttt{MismatchFlagHost} flag set to 1 (HD 106906, Kepler-420). 
These targets will be analyzed in Section \ref{sec:binaries}.

For all planets not showing issues with the  \texttt{binary} flag, the code performs a simple check on the coordinates, to find out if all entries for a single target are consistent with one another.

The code groups the various entries by the host star name. It then retrieves the \emph{mode} of the right ascension and declination (i.e. the value of each coordinate that appears more often in the group)  and checks if there are inconsistent values, that differ from the mode by more than 0.005 degrees. In that case, the wrong value is replaced by the mode of the coordinate itself.

If no mode is found (i.e. there is no most common value), no replacement is made: any inconsistency will be solved at a later point in the process. 
 The software sends a warning to the user, reporting that the four catalogs are not in agreement for what concerns either right ascension, declination, or both. This automated check helped us find errors within the original catalogs and warn the catalog maintainers about certain issues. 
 
 At the time of writing the code successfully found 200 inconsistent coordinates, most of which (about 110) replaced with the mode value. About two-thirds of these errors concerned the declination value. In some cases, especially for the lower values of declination (less than 1 degree),  a plus/minus sign difference appeared among the various datasets. This is caused by the inner uncertainties of such coordinates. \textit{Gaia} could improve accuracy by retrieving more precise coordinates and proper motions.

\subsection{Main identifier retrieval}\label{sec:mainid}

Despite all efforts made up to this point by the previous functions, in some cases, the host identifier for the same target could be different in the four catalogs, so any merging by host star name would still be inefficient. Besides, it could be useful to provide a link to the most important stellar catalogs for future analysis of the stellar-planet systems.

To accomplish this, the code performs a series of ADQL (Astronomical Data Query Language) queries to multiple TAP services such as SIMBAD and VizieR\footnote{http://vizier.u-strasbg.fr/viz-bin/VizieR} \citep{2000A&AS..143...23O}, to collect all useful data (in our case, identifier and coordinates) from the most important catalogs such as Kepler \citep{KIC} and K2/EPIC Input Catalogs \citep{K2}, as well as \textit{Gaia} DR2 \citep{GAIA,GAIADR2}.

First of all, the four DataFrames are concatenated to create a global DataFrame with more than 20000 entries belonging to the four catalogs. Indeed, we expect that the majority of such entries are duplicate datasets belonging to different catalogs.

The code loads the SIMBAD TAP service and queries it via \texttt{pyvo}. The first query looks for an exact match between the name of the host star as assigned in the global catalog, and the known identifiers in the SIMBAD Archive. All successful results from the query are stored in the corresponding \texttt{main\_id} and official coordinates \texttt{ra\_off}, \texttt{dec\_off} columns for all occurrences of each host star. 

For all host star names for which the exact string match was unsuccessful, a new query is made by considering all the known aliases contained in the \texttt{alias} column.

These queries are indeed effective in finding the appropriate main identifier for most of the targets: the number of missing main identifiers at this stage is reduced to about 400 entries, from an original number of more than 20400 elements (the concatenation of all four DataFrames). At this stage, all main identifiers found in the previous queries are unequivocally linked to the original denomination, being based on an exact match of strings. 

For all unsuccessful targets, another query to SIMBAD is then made by cross-matching the coordinates of each target with all sources within the online archive. These sources are considered to be potential matches with the corresponding target if their coordinates fall inside a circle of radius 0.0005 degrees from the coordinates provided by the considered exoplanet catalog. This value was chosen to account for the average precision of the right ascension and declination values that are available from the input catalogs.

In general, it may be possible that multiple sources are found within the circle, so the software calculates the angular separation from the original coordinates with \texttt{astropy}. Only the source with the shortest angular separation from the center is stored in the main identifier and default coordinates columns. 

In this case, all successful matches have very small angular separation and from a quick view it was possible to witness the fact that all identifiers linked correctly and the main identifier string was indeed similar to the original one, but the notation of the latter was unconventional and was not recognized by the previous query.

These steps sort out the vast majority of targets, leaving 175 entries in the general catalog still without a main identifier at the time of writing.

Switching the TAP service from SIMBAD to VizieR, the code can query the other catalogs by coordinates, in a similar way it previously did. The code queries the Kepler Input Catalog and the K2/EPIC Input Catalog since most of the known candidates are included in the Kepler surveys. In this way, only about 150 entries have still a missing identifier.

At this point, the software connects to the ARI-GAIA\footnote{http://gaia.ari.uni-heidelberg.de/} TAP service to query \textit{Gaia} DR2 Archive. This proves to be effective, leaving 94 targets with no identifier at tolerance 0.0005 degrees.

Since there are still targets without their main identifier, the code increases the tolerance of the query (i.e. the radius of the circle around the original coordinates) and tries the same queries again until all remaining items acquire the corresponding main identifier.

At tolerance 0.0025 degrees the queries to SIMBAD and K2/EPIC are still effective, leaving only 22 items with no identifier. At the same tolerance, GAIA finds 11 of them. Any further increase of tolerance seems to be effective only for GAIA DR2, which finds all other targets by a maximum tolerance of 0.0175 degrees.

From a manual check on these last 94 elements, for which the larger tolerance had increased the possibility of a mismatch, the correctness of every match was confirmed. 

At the time of writing, the current amount of targets in all source catalogs has correctly been taken care of. It is however impossible to exclude the need for some adjustment in the cross-match radius in the future, depending on new discoveries and their treatment in the original databases.

At this point, since the main identifier column allows to easily group all occurrences, we performed a check to find multiple entries of the same planet within the same source catalog. This was, unfortunately, the case for a few targets, mainly for the EU (73 duplicated entries), ORG (63 duplicated entries) and OEC (16 duplicated entries) catalogs. These planets were included in the catalog with both their provisional candidate name and with their confirmed one. Such issues are automatically identified by the software, and stored in a log file that could be sent to the catalog maintainers. The NASA Archive has no duplicated entries at all. 

\subsection{Catalog retrieval}\label{sec:catalog}

The cumulative catalog can be hierarchically indexed by the tuple \texttt{main\_id},  \texttt{binary} (if present) and \texttt{letter}. This is supposed to be more effective after the previous treatment on the homogeneity of notations. 

We expect to have up to four entries for each planet and the code has to collapse them to one single entry, based on the precision of the measurement. 

For each parameter (mass, minimum mass, radius, period, semi-major axis, inclination, eccentricity) the code calculates the relative error $X_{rel}$, defined as:

\begin{equation}
    X_{rel}=\frac{max(err^X_{min},err^X_{max})}{X}
\end{equation}

Where $X$ is the value of the considered parameter, while $err^X_{min}$ and $err^X_{max}$ the absolute values of the lower and upper error.

For every single parameter, the code selects the dataset (value and errors) with the smallest relative error, and it stores the reference paper in which the chosen dataset first appeared.

Until now, the column describing the planet name was left unchanged from its raw value, while the host star names were rearranged and standardized.
It could be then possible that multiple planetary names appear as default names for the same tuple \texttt{(main\_id, binary, letter)}. The code selects the string that contains the commonly known name, by privileging Kepler, WASP, Gliese, K2, HD, Hipparcos, CoRoT identifiers when available. The prioritization of some identifiers with respect to others is arbitrary and can be changed at will.

Aliases for the same target are stored in a cumulative list of strings.

At this point, each group of duplicates was collapsed into one single row. The fingerprint of the original group remains within the column \texttt{catalog}, which shows a list of all catalogs in which the item was found. This list could be then composed of four elements, if the planet was present in all input catalogs, rather than three, two or a single element. 

The status of the single target is stored in a string that retains information concerning the original label from the source catalogs. The strings follow the pattern \texttt{AXDXEXCX} where A represents the NASA \textbf{A}rchive, D the Exoplanet Orbit \textbf{D}atabase, E the Exoplanet \textbf{E}ncyclopaedia, and C the Open Exoplanet \textbf{C}atalog, while X is an integer from 0 to 3, where 0 means that the target is not present in the catalog (represented by the previous letter), 1 if it is labeled as a false positive, 2 if it is a candidate and 3 if it is confirmed.

To ease further analysis, the code also provides the most probable status of the planet: if the previous string is composed by 0 or 3 only (the planet may be not present in every catalog, but when present it is labeled as confirmed); if at least a 2 is present, the planet is labeled as candidate; if at least a 1 is present, the target is a false positive.

Once the final row for each target is ready to be concatenated to the rest of the output catalog, the code stores the values of the best mass, by choosing between mass and minimum mass the measurement with the smallest relative error. The fingerprint of the original measurement is stored in the column \texttt{mass\_prov}: it could, therefore, contain \texttt{mass} strings, if the most precise measurement was the mass, or \texttt{msini} strings otherwise.
This will be useful to plot any value of the mass, eventually choosing a different marker for minimum mass or mass of the planet (as described in Section \ref{sec:GUI}).

\subsection{Upload and Web Source}

The latest updated version of the catalog is made available as a VO resource through a dedicated TAP service. Subsequent versions of Exo-MerCat are generated as runs of the software described in this paper as a workflow. For a description of both the update runs and the VO deployment of Exo-MerCat refer to Sec.~\ref{sec:vo}.

\section{Performance}
\label{sec:performance}
In this section, we will try to assess the performance of the code, and in particular, we will be focused on the improvements in the final catalog that the various functions allow.

First of all, we merged the initial databases as they were downloaded from their sources. Of course, to be able to compare and sort the four datasets, they needed to be standardized, so that all interesting columns could have the same string as a header. No treatment whatsoever with filters, aliases, main identifiers, and coordinates was made.  

The function which retrieves the merged catalog was then executed, interpreting the \texttt{Host} column as the main identifier. In the remainder of the section, we will refer to this particular run as a "Simple" among the catalogs.

We then performed a full run of the Exo-MerCat software, executing all of the aforementioned functions (status, alias, and coordinates checks, as well as the main identifier retrieval). We will refer to this run as an Exo-MerCat (in short, "EMC") match among the catalogs.

Therefore, we performed twice the same merging of the four catalogs: while the first time the original state of the four archives was preserved, the second run showed the full potential of the software. By comparing the results, we expect to be able to analyze the improvements that this code allows us to achieve.

\begin{figure*}
 \centering
 \includegraphics[width=\linewidth]{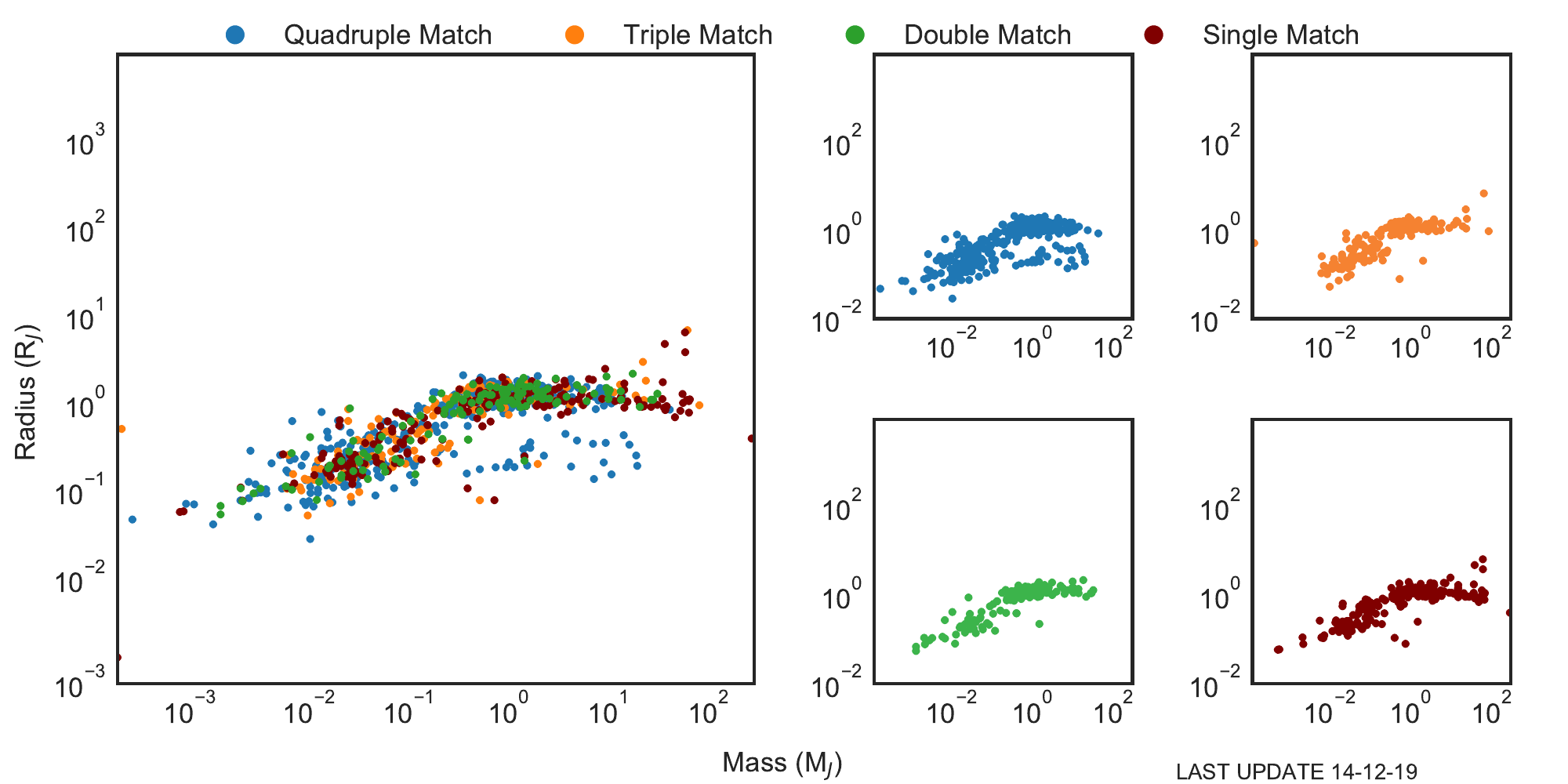}
 \caption{Mass-Radius plot for the Simple Run. As shown in the legend, blue dots refer to all quadruple matches (i.e. planets found in all databases), the orange dots to the triple matches (i.e. planets found in three over four databases), the green dots to the double matches (i.e. planets found in two over four databases), and the red ones to the single planets (i.e. planets found only in one database). }
 \label{figure:MRnocode}
\end{figure*}

\begin{figure*}
 \centering
 \includegraphics[width=\linewidth]{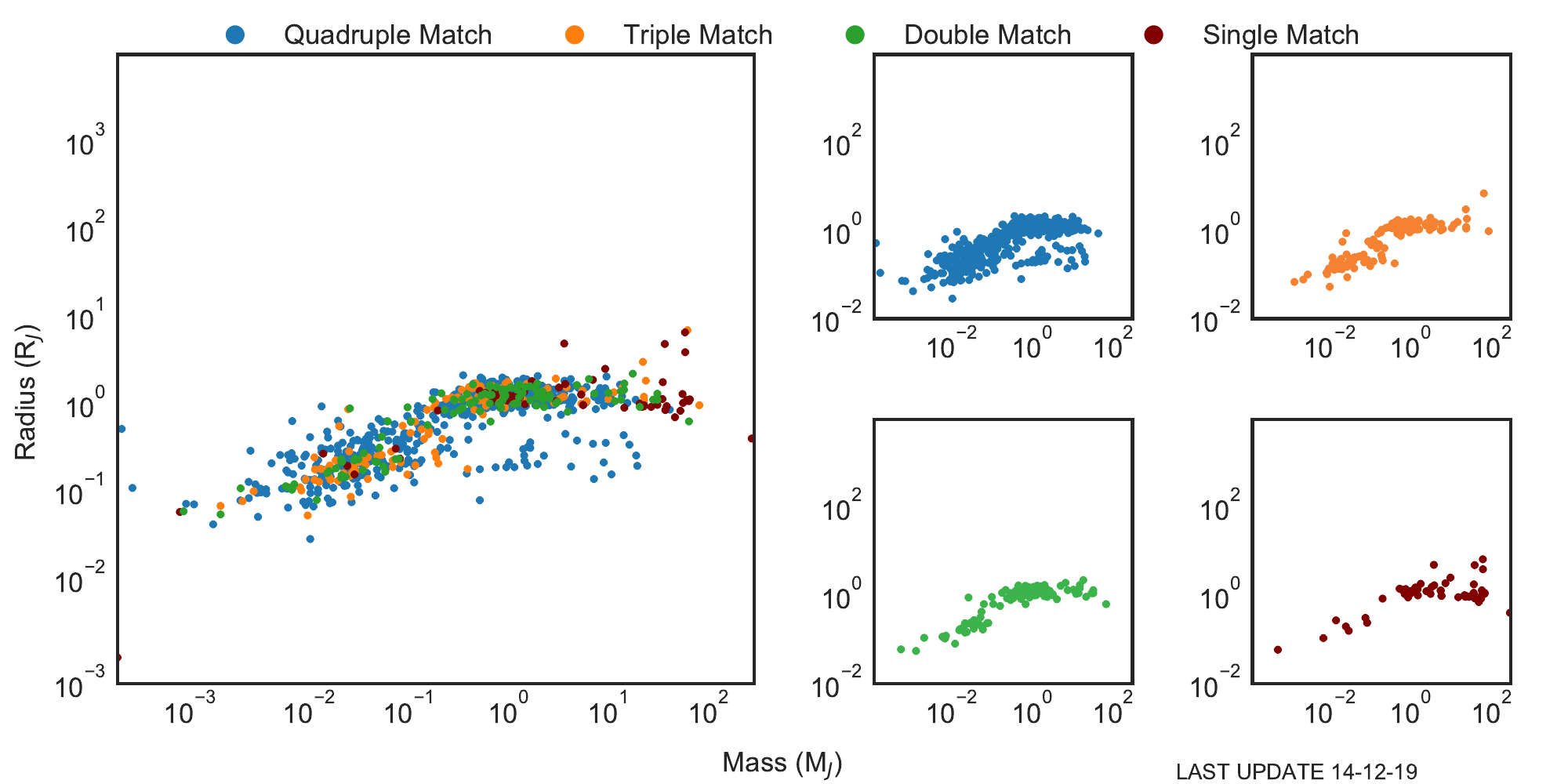}
 \caption{Mass-Radius plot for the EMC run. As shown in the legend, blue dots refer to all quadruple matches (i.e. planets found in all databases), the orange dots to the triple matches (i.e. planets found in three over four databases), the green dots to the double matches (i.e. planets found in two over four databases), and the red ones to the single planets (i.e. planets found only in one database). }
 \label{figure:MRcode}
\end{figure*}

The results are shown in Table \ref{tab:codenocode}. The M-R plot with all quadruple, triple, double and single matches is shown in Figures \ref{figure:MRnocode} and \ref{figure:MRcode}.

\begin{table*}
  \centering
  \caption{Results of the merged catalog with and without corrections.   Update: December 14, 2019. }
 \label{tab:codenocode}
\begin{tabular}{ l  r  c c }
\hline\hline

Samples && EMC RUN & SIMPLE RUN \\\hline
 \multicolumn{2}{c}{All  Planets} & 7538 & 10300\\
\multicolumn{2}{c}{All Confirmed/Candidate Planets} &93.6\%& 99.8\%\\
\multicolumn{2}{c}{Quadruple Matches} & 42.4\%&28.0\%\\
&\small\textsc{Confirmed}&\small3133& \small2855\\
&\small\textsc{Candidate}&\small61 &\small 27\\
&\small\textsc{False Positive}&\small2 &\small0 \\
\multicolumn{2}{c}{Triple Matches} &5.2\%& 4.8\%\\
&\small\textsc{Confirmed}&\small335 &\small477 \\
&\small\textsc{Candidate}&\small57 &\small24 \\
&\small\textsc{False Positive}&\small4 &\small1 \\
\multicolumn{2}{c}{Double Matches} & 31.6\%& 5.5\%\\
&\small\textsc{Confirmed}&\small416 & \small517\\
&\small\textsc{Candidate}&\small 1962 & \small 51 \\
&\small\textsc{False Positive}&\small6 &\small3\\
\multicolumn{2}{c}{Single Match} & 20.8\%& 61.7\%\\
&\small\textsc{Confirmed}&\small375 &\small 1121\\
&\small\textsc{Candidate}&\small718 &\small 5208 \\
&\small\textsc{False Positive}&\small469 &\small 16\\

\hline
\end{tabular}

\end{table*}


First of all, Table \ref{tab:codenocode} shows that the Simple run causes the whole catalog (confirmed planets, candidates, and false positives) to be composed of 10300 elements. On the other hand, for the EMC run the final amount of planets is about 7500 (nearly 4200 confirmed targets, 2800 candidates, and 480 false positives). This means that the software indeed is effective in finding a large number of duplicates. 

For the Simple run, nearly half of the sample is present in the Single Match subgroup. This does not happen for the EMC run, which on the other hand shows nearly half of the same amount of candidates in the Double Matches subgroup: this means that nearly 3000 duplicate/multiple candidates were indeed present in the sample, but were not recognized in the Simple run because of the difference in their notation -- and were therefore categorized as single occurrences. 
Most of the items of this subgroup belonged either to the EU or the ORG archive (which at present are the only two catalogs that provide a substantial amount of candidates by default) and in those cases the notation used for the KOI objects is different (KOI-NNNN.DD for the ORG, KNNNN.DD for the EU): this was the cause of the low efficiency in the match, which on the other hand can be corrected by the Exo-MerCat catalog. 

The amount of quadruple matches for the two cases shows a difference of nearly 300 planets, giving us more confidence concerning the effectiveness of the software. The presence of 60 candidates in this subgroup shows that there is a slight amount of candidates which is common for all catalogs, maybe due to a lack of updates. 

By comparing the M-R plots (Figures \ref{figure:MRnocode} and \ref{figure:MRcode}), we notice a much smaller amount of data contained in the subgroup of planets appearing in only one catalog (red sample) for the EMC run with respect to the Simple run. On the other hand, the subgroup of planets appearing in all catalogs (blue sample) appears to contain more data. This depends both on the effectiveness of the match for the various targets, but also on the data selection, which allows the final catalog to have, in the end, a higher number of measurements belonging to different reference papers. 
The trend determined by the theoretical relation present in the ORG catalog is a priori removed in the Exo-MerCat software.

\begin{figure*}
 \centering
 \includegraphics[width=0.7\linewidth]{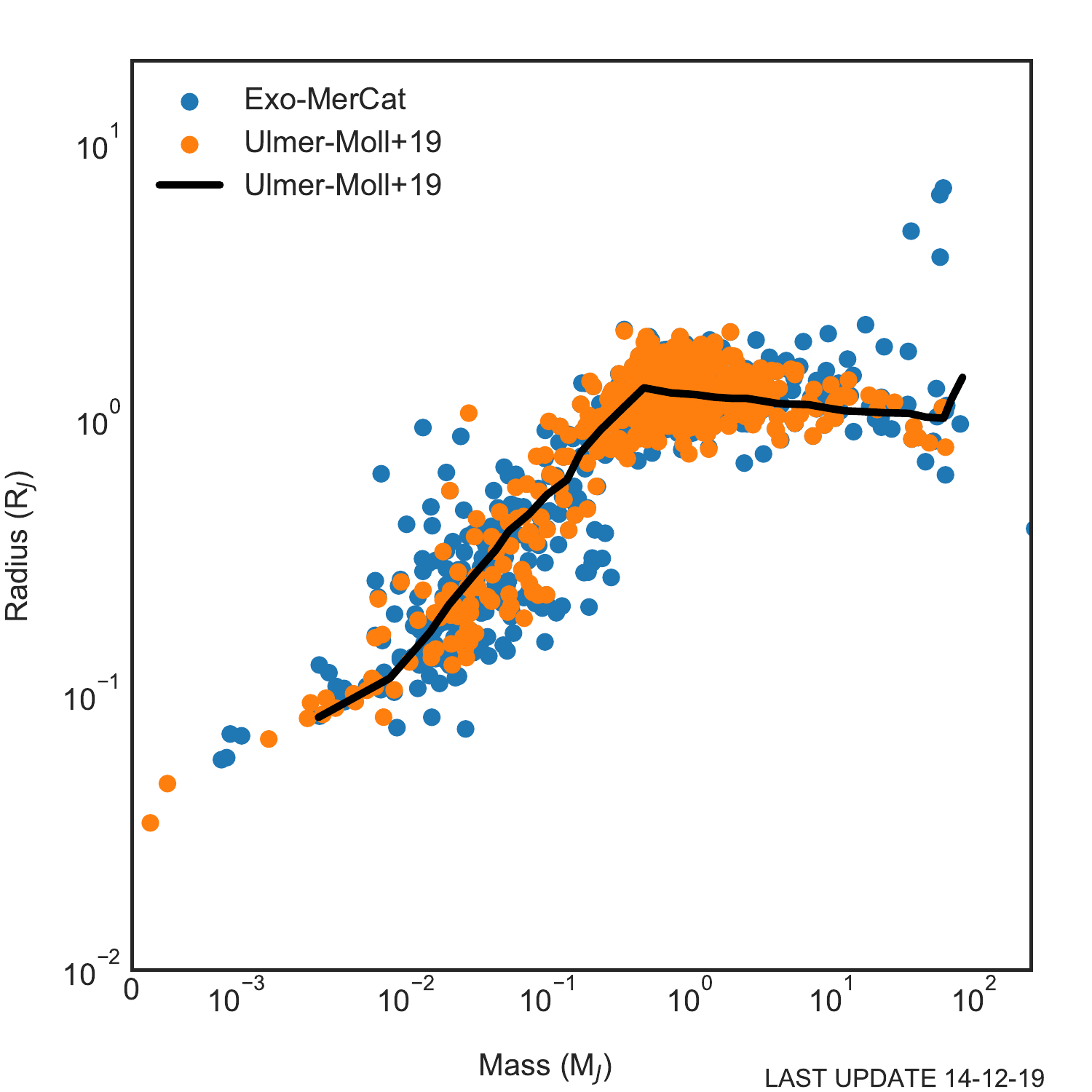}
 \caption{Mass-Radius plot for the validation sample from \citet{UlmerMoll2019} (orange dots) and the sample of planets whose relative error on mass and radius is smaller than 80\% produced by Exo-MerCat (blue dots). In black, the theoretical mass-radius relation calculated in \citep{UlmerMoll2019}. }
 \label{figure:MRUlmerMoll}
\end{figure*}

We then compared the available measurements of mass and radius contained in Exo-MerCat with the verification sample used in \citet{UlmerMoll2019}, as shown in Figure \ref{figure:MRUlmerMoll}. 

The validation sample considered by the authors is composed of 506 objects with mass and radius measurements, as retrieved from the EU catalog in April 2019. The authors built a random forest algorithm trained and validated on the verification sample to estimate the radius of exoplanets based on their mass, obtaining relationships similar to the black solid line in Figure \ref{figure:MRUlmerMoll}.

The sample produced by Exo-MerCat and shown in Figure \ref{figure:MRUlmerMoll} is composed uniquely by the measurements whose relative errors on mass and radius measurements were smaller than the 80\%, for a total of 758 elements.

Unsurprisingly, the theoretical relation appears to be in agreement with both the data belonging to Exo-MerCat, as well as the validation sample. The sample produced by Exo-MerCat, however, contains more elements, with a few candidates covering regions in the mass-radius parameter space which are not included in the verification sample used by \citet{UlmerMoll2019}. It could be therefore possible that, when repeating similar analyses, small differences in the theoretical relations would appear. 

\section{Binary Host Stars}
\label{sec:binaries}

As stated before, it is impossible to figure out a complete sample of binary host stars due to the lack of information from the original catalogs. This is caused by a well-known ambiguity in the nomenclature of such host stars since the discovery of binary companions can often be tricky. Sometimes, most of the primary companions are already known in the community and a change in the notation due to the subsequent discovery of a companion would be confusing. 

Other times, the two stars are already known with different names well before discovering that they are gravitationally bound, so the names cannot vary to avoid the chance of mismatch in the literature.

For these and many other reasons, it is no wonder that even the exoplanet databases suffer from such discrepancy. For our purposes, however, this issue does not preclude the possibility of correctly comparing the various targets in the vast majority of cases, since it would suffice to create uniform strings to be compared. 

We managed to do that in the Exo-MerCat code, as stated previously. However, it is essential to keep in mind that the current version of the code cannot provide a complete sample of all planets orbiting one or more binary companions, since the value of the \texttt{binary} string is not indicative of the effective architecture of the system in most cases, but it is only useful to allow the software to correctly compare planets belonging to complex systems. 

At the time of writing, the final catalog had 191 non-null values of the \texttt{binary} column: within this sample, 68 planets were circumbinary. The unique host star identifiers in the binary sample were 143. The targets displaying a non-null \texttt{MismatchFlagHost} flag were 2: therefore, the software failed in recognizing any duplicate entries among these targets. 

The remainder of the section will be devoted to a more detailed study of these targets.

\begin{itemize}
    \item HD 106906 b is a planet discovered by the imaging technique, found in NASA, OEC, and EU catalogs. In the OEC Database, this planet is labeled as circumbinary, while there is no information concerning the architecture of the planet in the other archives. When checking the coordinates for all entries in the global DataFrame, a difference in coordinates up to 2 degrees appears in the EU catalog and this forbids any correction for this row. The coordinates stored in NASA and OEC archives are, on the other hand, in good agreement and for this reason, the correction of the \texttt{binary} label can be made for this couple. In the final catalog, therefore, two rows for this planet, with different coordinates and different values of the \texttt{binary} string.  In any case, SIMBAD successfully retrieves the correct coordinates since the host star name is well known within the archive. The ambiguity between the catalogs is probably due to the recent discovery of the binary nature of HD 106906, which belongs to the Lower Centaurus Crux (LCC) group in the ScoCen association, which was claimed a few years after the discovery of the planetary companion \citep{Bailey2013}. The latter is on a wide orbit (with a period of about 3000 years), while the binary stars are very close, with an orbital period of 100 days \citep{Rodet2017}, so it is highly probable that the planet is on a P-type orbit. For this reason, we forced the merging of the two duplicated rows, discarding the set of coordinates belonging to the Exoplanet Encyclopaedia and setting the \texttt{binary} string to "AB".
    
    \item Kepler-420 b is a transiting planet discovered with Kepler and confirmed with follow-up radial velocities observations. This target is present in the four catalogs, with different names: in the EU and ORG catalog, the original string was either Kepler-420 A b or Kepler-420 b; in NASA and OEC catalogs, the KOI notation was present (KOI-1257 b). This target had a non-null binary flag in all source DBs except for the NASA Archive.  
    
    As stated by \citet{Santerne2014}, this target is a planet that is most likely to orbit the primary companion of a binary/multiple group, so the expected \texttt{binary} string should be A. 
    The check in coordinates is not effective in this case to identify the four entries as the same target, since the value of the declination is about half a degree different in the EU catalog with respect to the other ones, despite being the other parameters consistent in all the archives. The main identifier is however easily found in SIMBAD. 
    
    We, therefore, had to simply force the value of the \texttt{binary} string to be A for all entries to ensure a perfect match among the various entries.
    
\end{itemize}

During the tests and the validation of the software, also ROXs-42 B b had a non-null potential mismatch flag. This is a planet on a circumbinary orbit around the binary ROXs-42 B (AB) in the Ophiucus Molecular Cloud. This target is found in NASA, EU, and OEC archives. For the latter, the code managed to retrieve information about the P-type orbit, while no information is retained in the remaining two. The software, at first, interpreted the B in the planet name string and assigned the value "B" to the \texttt{binary} cell. When comparing the various entries during the coordinate check, for this target a disagreement between two non-null values of  \texttt{binary} was present. In this case, the issue originated from the very name of the stellar system, which is labeled B not because of its binary nature, but because it is the second-brightest counterpart of the X-ray source ROXs-42. This source is too largely separated from ROXs-42 A and C to be gravitationally bound to those \citep{Kraus2013}, but it is a close binary system by itself \citep{Simon1995}. 

Further updates on the original databases automatically solved this issue, allowing all available entries to be in agreement concerning the nature of this planet-star system.

It is worth to stress again that we do not expect to find a complete sample of all known planets orbiting one or more stars in a binary/multiple system with the available information. Further studies and checks performed with the known binary catalogs are due and will be explored in future publications. A comparison is foreseen with the webpage\footnote{http://exoplanet.eu/planets\_binary/} dedicated to the known planets in binary systems within 200 au in the Exoplanet Encyclopaedia.

\section{Brown Dwarfs}
\label{sec:bd}

As shown in Table\ \ref{tab:criteria} and \ref{tab:stats}, the four original databases do not seem to have similar constraints over the mass and radius values of the objects to include in the sample. Sometimes, the values that are present in the database do not follow the selection criteria reported in the official documentation. This could be surely caused by errors or lack of an updated version of either the database itself or the ancillary documentation files, but it underlines an issue that is still debated, concerning the nature of the brown dwarf sample. 

 These are intermediate objects with masses so low to cause the electron degeneracy pressure to forbid hydrogen burning in their cores, which is an essential requirement to be a low-mass star. For a review on these objects see e.g. \citet{luhman2012}. Because of this process, the upper boundary for a brown dwarf's mass is supposed to be less than 72 $M_{Jup}$ (the hydrogen-burning mass limit, see e.g. \citet{Schneider2018}). 
 
 Earlier in their evolution, brown dwarfs can burn deuterium (or lithium, for more massive objects -- about 60 $M_{Jup}$). This process should determine the lower boundary of $\approx13 \ M_{Jup}$ as the threshold which divides brown dwarfs from massive exoplanets.  This boundary is however not clear, as it may depend on the initial helium and deuterium abundances, as well as the metallicity of the stellar model \citep{Spiegel2011}.
 
Formation by itself should provide a boundary between planets and brown dwarfs: the first class of objects is expected to gain mass by accretion of planetesimals from the surrounding dust disk, while the other is expected to form by gravitational collapse from the original gas cloud. 

However, formation models still provide a great degeneracy on the expected mass of both classes of objects, depending on the initial condition of the system which is no longer observable: indeed, different formation processes could lead to higher or smaller masses, making the 13 $M_{Jup}$ boundary less relevant.

Since the brown dwarfs cool progressively with time and observables such as temperature, luminosity, and mass, they often overlap with young massive exoplanets, or extremely old small stars \citep{Faherty2018}.

Because of this degeneracy in the observables, the classification of such objects is somewhat still arbitrary. 

The Exoplanet Encyclopaedia, for example, follows the arguments supported by \citet{Hatzes2015}: the mass-density relation follows a well-defined trend up to 60 $M_{Jup}$, but after that, a dramatic change in the slope happens, leading to much smaller densities and larger masses, typical for stellar objects. This theory, however, relies on very few observed objects in the 30/60 $M_{Jup}$ (the so-called brown dwarf desert), as well as the difficulties in measuring radii for the available objects \citep{Schneider2018}. Large errors on the mass measurement are also probable, especially for direct imaging candidates, since they rely only on photometry and models \citep{Schneider2018}.

Major improvements on this topic are foreseen with future radial velocity surveys as well as astrometric data from \textit{Gaia}.

\begin{figure}
 \centering
 \includegraphics[width=0.8\textwidth]{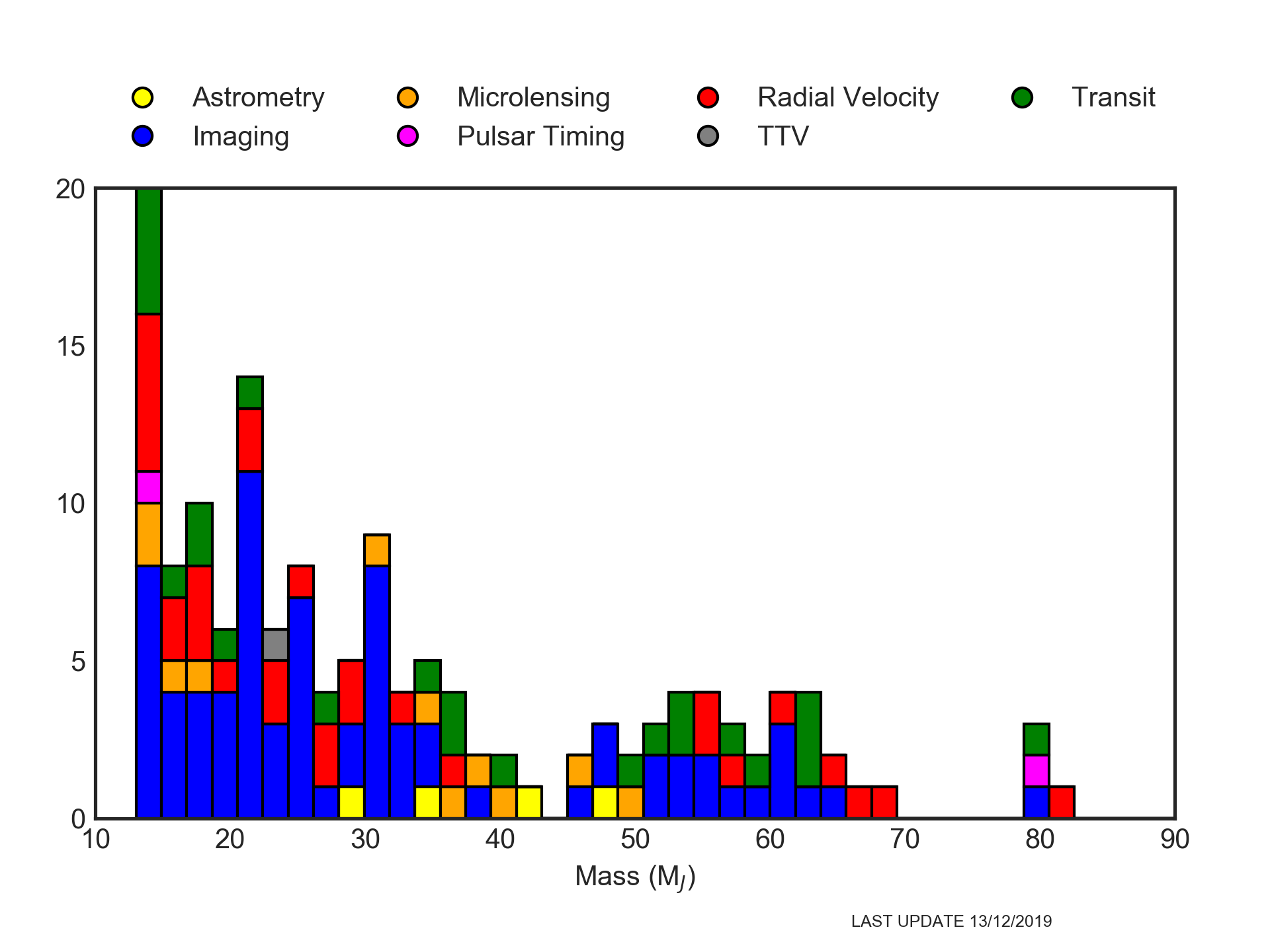}
 \caption{Distribution of the best masses of brown dwarf sample.  }
 \label{figure:bdhist}
\end{figure}

At the time of writing, 206 targets whose mass is higher than 13 $M_{Jup}$ are present in the merged catalog (see Figure\ \ref{figure:bdhist}). Most of them belong to the Exoplanet Encyclopedia and are assumed as confirmed (175), while 29 are labeled as candidates and 2 as false positives. 

The sample of objects whose mass is higher than 60 $M_{Jup}$ reduces to 17 targets, the majority of which are objects at slightly higher masses with large enough errors to be under the boundary. The two false positives belong to this sample as well, four are candidates and the remaining ones are confirmed targets. In this case as well, nearly all objects belong to the Exoplanet Encyclopaedia.

\section{Catalog update and VO interoperable access}
\label{sec:vo}

The catalog described in this paper is both susceptible to updates (due to changes in the originating catalogs and improvements in the knowledge of exoplanets) and useful when used in combination with other astrophysical databases and information. The former characteristic makes it clear that subsequent runs of the code that generates it are needed, the latter brings in the interoperability scenario and thus the idea of having it exposed as a VO resource.

\subsection{Exo-MerCat upate workflow}
\label{sec:workflow}

In Section \ref{sec:exomercat} we showed the software to merge the four main exoplanet catalogs available online.
According to Figure \ref{figure:flow}, the workflow for a single execution of Exo-MerCat is made up of multiple phases, executed step by step to get a final merged consistent catalog.
Moreover, source catalogs are periodically updated, therefore we need to 
run the Exo-MerCat procedure to update the merged catalog accordingly.

We managed to describe the Exo-MerCat workflow with the Common Workflow Language (CWL)\footnote{https://www.commonwl.org}, a standard specification designed purposely for describing analysis workflows and tools with YAML (YAML Ain't a Markup Language) structured text files. This choice comes with several advantages.
Since CWL is a standard, it can provide a high level of interoperability and portability across different hardware environments. CWL is excellent for jobs that must be run periodically, as Exo-MerCat, because all input parameters are stored in YAML files, which can be versioned. Besides, it is possible to keep track of the execution history with such YAML files, opposite to the execution of individual command lines, or at least it is more complex. Finally, the wide flexibility of CWL in workflow description is optimal to keep the whole workflow description updated for any future upgrade of Exo-MerCat software.

In the last phase, the final merged catalog is ingested 
into a proper database with TAP service as described in the next Section \ref{sec:emctap}.
At present, the Exo-MerCat workflow is configured to be executed once a week
to integrate updates of the four source catalogs. 
The total execution time of the weekly workflow is on average 25 minutes.

\subsection{Exo-MerCat VO resource}
\label{sec:emctap}

Once Exo-MerCat is available as a table in a database system (a step taken care of as described in the previous subsection, \ref{sec:workflow}), it becomes easy to annotate it with proper metadata, like the description and other details available from Table~\ref{tab:catdesc}, and register it as a resource in the VO ecosystem.

What has been done for Exo Mer-Cat has been to:
\begin{itemize}
    \item include and describe the catalog table within a TAP service;
    \item register the catalog as a VO resource;
    \item (register the above TAP service).
\end{itemize}

The technical details for the first two points above will be given in Appendix (\ref{app:vodetails}).
The third item is in parenthesis because the mentioned deployed TAP service will serve multiple data resources including, but not limited to, the Exo Mer-Cat catalog. This means its registration details are outside the scope of this description.

\begin{figure*}
 \centering
 \includegraphics[width=\linewidth]{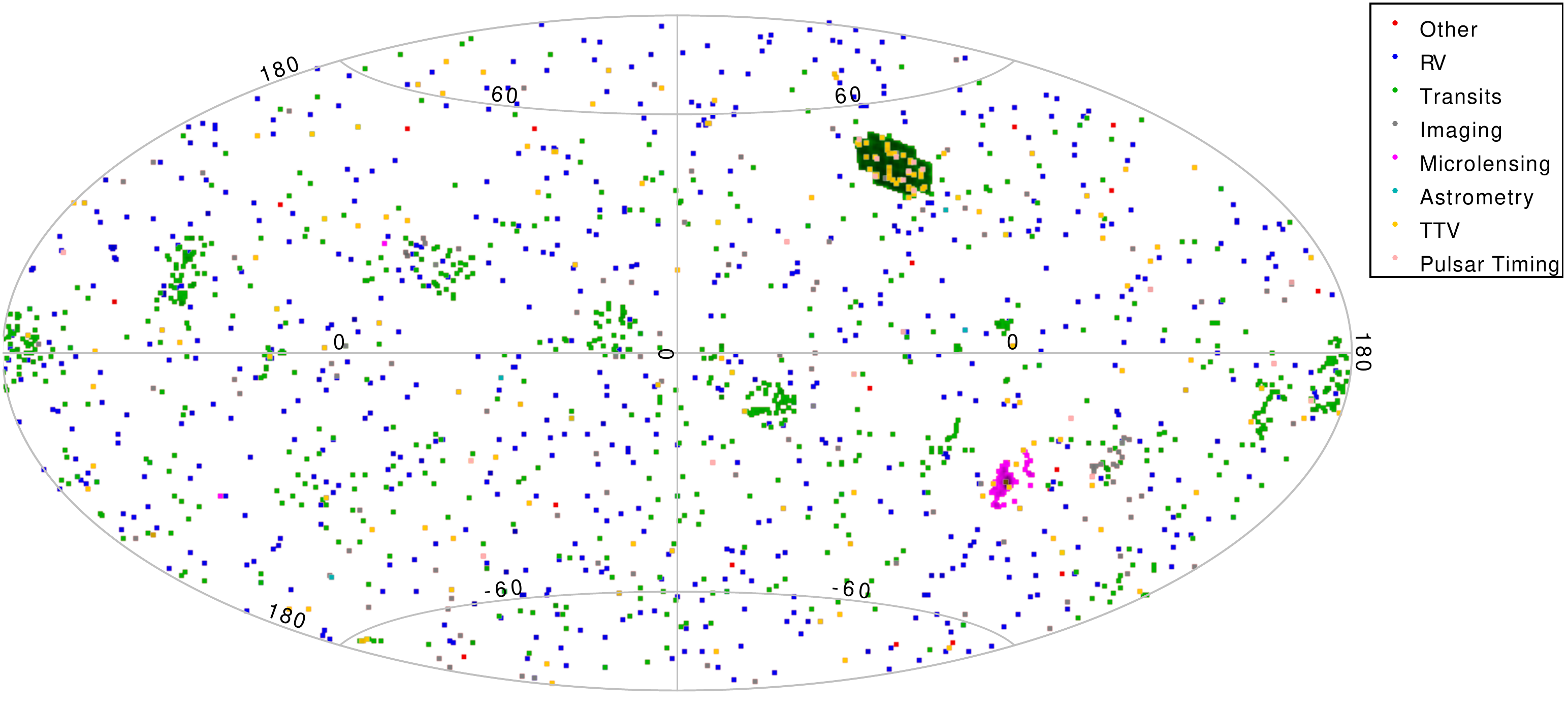}
 \caption{Distribution of the coordinates of  Exo-MerCat objects (color-coded depending on the discovery method) as shown by TOPCAT (Aitoff projection of equatorial coordinate system).}
 \label{figure:topcat}
\end{figure*}

Such metadata description and resource registration allow for the catalog to be visible and consumable by all the VO-aware TAP-enabled client applications, like TOPCAT \citep{2005ASPC..347...29T}. This will improve catalog visibility and interoperability of the data resource. As an example of the use of TOPCAT, in Figure\ \ref{figure:topcat} is shown the distribution of the objects contained in the Exo-MerCat catalog in the Galaxy. The knob of {\it Kepler} objects is easily recognizable.

Currently, the TAP service is available through the URL\footnote{http://archives.ia2.inaf.it/vo/tap/projects} in the footnote, but it is recommended to use the details and identifiers reported in the appendix to connect to the service to be sure to reach the proper resource.

\section{Graphical User Interface}
\label{sec:GUI}
\begin{figure*}
 \centering
 \includegraphics[width=\linewidth]{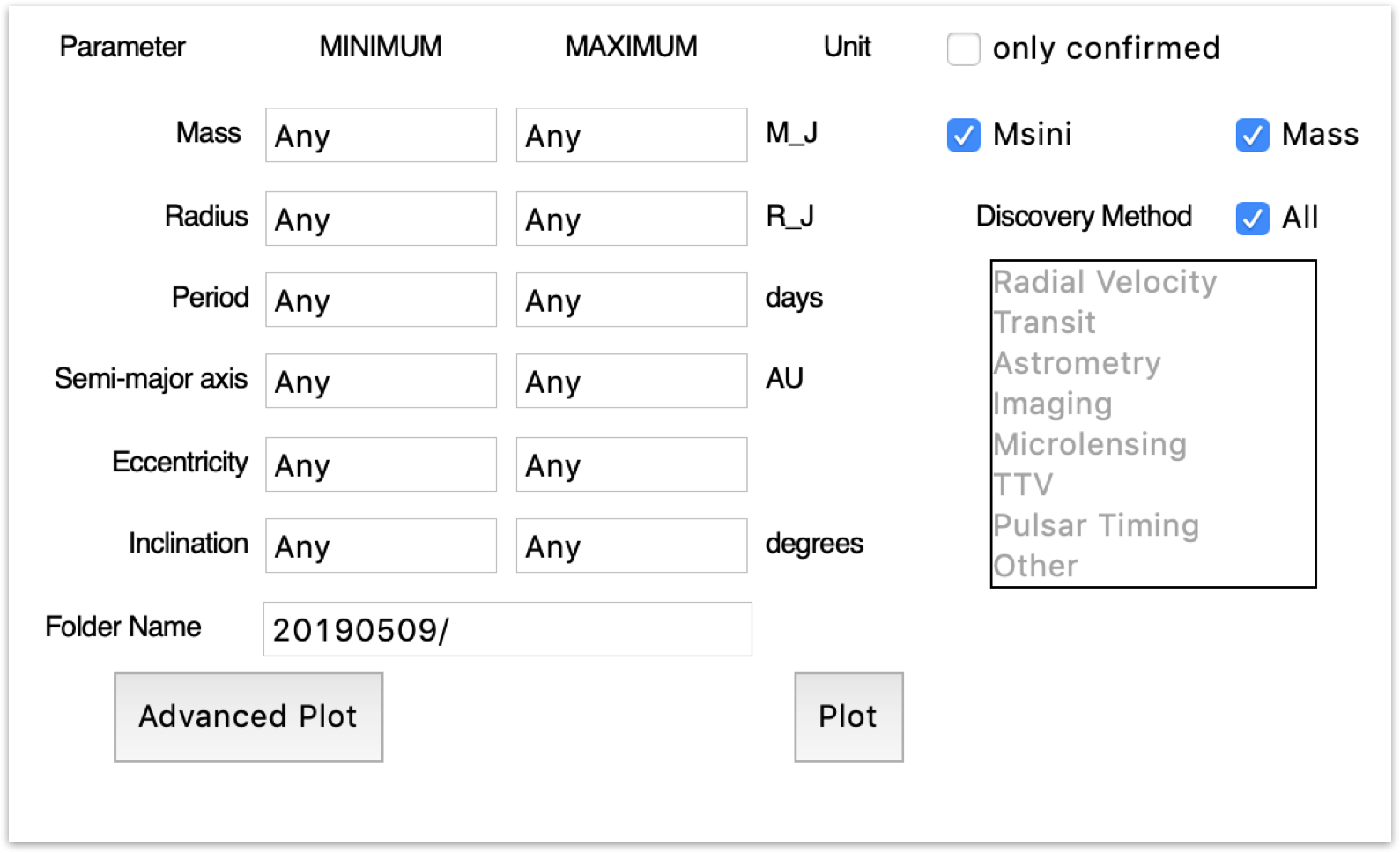}
 \caption{Screenshot of the Graphical User Interface that filters and plots data from the Exo-MerCat catalog.}
 \label{fig:GUI}
\end{figure*}

The Graphical User Interface can be downloaded from a public GitHub repository\footnote{https://gitlab.com/eleonoraalei/exo-mercat-gui}. An image of the GUI at its current state is shown in Figure \ref{fig:GUI}. 

It is a Python 3.6 script, which requires \texttt{pyvo} to query the Exo-MerCat TAP service. In addition to that, \texttt{matplotlib}, \texttt{numpy}, \texttt{pandas}, and \texttt{guizero} need to be installed. This can be easily done by installing via command line using the provided file \texttt{requirements.txt}.

The package \texttt{guizero}\footnote{https://lawsie.github.io/guizero/about/} allows creating a highly customizable interface that offers the user the option to filter the catalog. The user can choose the upper or lower limit for any of the following parameters: mass, radius, period, semi-major axis, eccentricity, and inclination. It is furthermore possible to exclude candidates and false positives and to select one or both mass measurements (mass and minimum mass), if available. 
It is also possible to select all discovery methods or just a few.

By default, the filtered catalog and the plots will be stored in a folder named as the current date in format \texttt{YYYYMMDD}. If not already present, the folder is created by the GUI code itself. The user can, however, change the name of the folder. The path can be changed by specifying its relative position with respect to the enclosing folder (i.e. the folder where the GUI script is saved).

The script checks if the threshold values the user provides are correct and adequate. In particular, when trying to press any of the two buttons at the bottom of the interface, a function is called to check whether a minimum is greater than the corresponding maximum, and/or if a value is set as a nonphysical negative, for each of the selected parameters. If at least one error appears, an alert window is displayed. Any incorrect text is colored in red. The user can correct each value until everything appears to be correct. If this is the case, an infobox pops up with a summary of all selected filters.

The values are stored in the corresponding variables, which are then used to filter the catalog.

By clicking the PLOT button, the script checks the threshold values, filters the catalog and automatically produces a set of standard plots, which are stored in the chosen folder:

\begin{itemize}
\item \emph{Radius ($R_\oplus$) vs. Best Mass ($M_\oplus$)}
\item \emph{Distance ($pc$) vs. Best Mass ($M_\oplus$)}
\item \emph{Period ($days$) vs. Best Mass ($M_\oplus$)}
\item \emph{Semi-major axis ($au$) vs. Best Mass ($M_\oplus$)}
\item \emph{Eccentricity vs. Best Mass ($M_\oplus$)}
\item \emph{Eccentricity vs. Semi-major axis ($au$)}
\end{itemize}

\vspace{1em}

And histograms: \emph {Best Mass ($M_\oplus$)}, \emph{Radius ($R_\oplus$)}, \emph{Semi-major axis ($au$)}, \emph{Period (days)}, \emph{Eccentricity}, \emph{Inclination ($deg$)}.

The script sorts data depending on the discovery method of each target, thus displaying the items in different colors. A string is also added on the bottom right corner of each plot to show the latest update of the graphs. 

If both mass measurements are selected, the plot shows the value of the best mass measurement i.e. the one with the lowest relative error. If the best mass is the minimum mass itself, the target marker is a diamond instead of a circle.
For the mass histogram, the best mass is the plotted parameter.

By default, all axes in plots are displayed in logarithmic scale, except those plots concerning the eccentricity. Also, all error bars are shown. All axes in the histograms are also displayed in logarithmic scale, except the horizontal axes concerning eccentricity and inclination.

However, depending on the filtering, it may be useful to switch on and off the logarithmic scale for any of the axes. This can be done by clicking the ADVANCED PLOT button. 
In this case, a second panel is opened to show a set of checkboxes (the default values will appear as already selected). The user can deselect an entire plot, so that it won't be created, and/or determine the scale of each axis. For greater clarity, error bars can be deselected.

\section{Conclusions}
\label{sec:conclusions}

We presented Exo-MerCat, a new tool to create a coherent catalog of exoplanets by comparing and matching the datasets available in the most important online catalogs. The catalog is available for all VO-aware TAP-enabled client applications and it is periodically updated. It relies on the usage of VO tools and standards, from a perspective of the more and more common usage of such technologies in the future, to ease datasets availability, maintenance, and coherent analysis. 

The effort is still ongoing with further improvements and the development of new features, such as the possibility to query for one or more older versions of the catalog: this is essential to allow any astronomer to dig deeper into the history of a certain target, by studying the variation of any measurement in time; it could be furthermore useful to retrieve an old version of the catalog, corresponding to the sample of targets known up to a specific date in time, to compare it with the current sample. 

Other catalogs may be connected to this one, possibly linking the targets to the available observed data, whether raw or already refined by some data analysis. A more automated connection with the host star catalogs could be also established, to allow the user to retrieve useful information concerning the star.

We should point out that this script would be nothing more than a cross-match among different sources if only the currently available exoplanet catalogs were fully VO-aware, with a common Unified Content Descriptor (UCD) for each parameter. This process of database standardization is starting to be more and more common for the most important stellar catalogs, which can now be easily handled by any VO-aware tool. 

Due to the youth of this field, this standardization is still not so straightforward for the exoplanets. For this reason, we started to work (under the EU H2020 ASTERICS project) on the delineation of a specific Data Model for this class of targets, which will pick up model components from the IVOA specifications and attach new ones when needed. 
We expect that, soon, this new Data Model would be fully integrated into Exo-MerCat, and that many other sources would choose to follow the path towards standardized labeling of all planetary-related parameters.

\section*{Acknowledgements}

The authors thank the referees for their insightful comments, which helped to improve this work. 

This work has made use of data from the European Space Agency (ESA) mission
{\it Gaia}\footnote{https://www.cosmos.esa.int/gaia}, processed by the {\it Gaia}
Data Processing and Analysis Consortium (DPAC\footnote{https://www.cosmos.esa.int/web/gaia/dpac/consortium}). Funding for the DPAC
has been provided by national institutions, in particular, the institutions
participating in the {\it Gaia} Multilateral Agreement.

This research has made use of the SIMBAD database, operated at CDS, Strasbourg, France.

This research has made use of the VizieR catalog access tool, CDS, Strasbourg, France (DOI: 10.26093/cds/vizier).

The workflow system to manage the catalog and the VO deployment have been provided using 
the resources and tools made available by the Italian center for Astronomical Archives (IA2),
part of the Italian National Institute for Astrophysics (INAF).

The authors thank Cyril Chauvin from the Exoplanet Encylopaedia team and Jason Wright from the Exoplanet Orbit Database team for the support.

\appendix
\section{Catalog headers}
\label{app:cathead}

Table~\ref{tab:catdesc} reports the columns the catalog is composed of, with descriptions and data type domain. This forms the basis from which metadata information has been derived for the inclusion of the catalog
as a VO resource (see App.~\ref{app:vodetails}).

  

\begin{longtable}{l|p{6cm}|l}
\caption{Default column headers, meaning, and type.\label{tab:catdesc}}\\
  \hline\hline
  Header & Meaning & Type\\\hline
name& The name of the planet. & STRING\\
host& The name of the host star.& STRING\\
letter&  The letter labeling the planet.& STRING\\
mass&  The mass of the planet in Jovian masses.& FLOAT\\
mass\_max& The positive error on the mass measurement in Jovian masses.& FLOAT\\
mass\_min& The negative error on the mass measurement in Jovian masses.& FLOAT\\
mass\_url&The bibcode of the reference paper in which the mass value first appeared.& STRING\\
msini& The minimum mass of the planet in Jovian masses.& FLOAT\\
msini\_max& The positive error on the minimum mass measurement in Jovian & FLOAT\\
msini\_min& The negative error on the minimum mass measurement in Jovian masses.& FLOAT\\
msini\_url& The bibcode of the reference paper in which the minimum mass value first appeared.& STRING\\
bestmass& The most precise value between mass and minimum mass of the planet in Jovian masses.& FLOAT\\
bestmass\_max& The positive error on the best mass measurement in Jovian masses.& FLOAT\\
bestmass\_min& The negative error on the best mass measurement in Jovian masses.& FLOAT\\
bestmass\_url& The bibcode of the reference paper in which the mass/minimum mass value first appeared.& STRING\\
mass\_prov&  A string labeling if the Best Mass is the mass itself, or the minimum mass.& STRING\\
p& The period of the planet in days.& FLOAT\\
p\_max& The positive error on the period measurement in days.& FLOAT\\
p\_min&The negative error on the period measurement in days.& FLOAT\\
p\_url& The bibcode of the reference paper in which the period value first appeared.& STRING\\
r& The radius of the planet in Jovian radii.& FLOAT\\
r\_max& The positive error on the radius measurement in Jovian radii.& FLOAT\\
r\_min& The negative error on the radius measurement in Jovian radii.& FLOAT\\
r\_url& The bibcode of the reference paper in which the radius value first appeared.& STRING\\
a& The semi-major axis of the planet in au.& FLOAT\\
a\_max& The positive error on the semi-major axis measurement in au& FLOAT.\\
a\_min& The negative error on the semi-major axis measurement in Jovian masses.& FLOAT\\
a\_url& The bibcode of the reference paper in which the semi-major axis value first appeared.& STRING\\
e& The eccentricity of the planet (between 0 and 1).& FLOAT\\
e\_max& The positive error on the eccentricity measurement in Jovian masses.& FLOAT\\
e\_min& The negative error on the eccentricity measurement in Jovian masses.& FLOAT\\
e\_url& The bibcode of the reference paper in which the eccentricity value first appeared.& STRING\\
i& The inclination of the planet in degrees.& FLOAT\\
i\_max& The positive error on the inclination measurement in degrees.& FLOAT\\
i\_min& The negative error on the minimum mass measurement in degrees.& FLOAT\\
i\_url& The bibcode of the reference paper in which the inclination value first appeared.& STRING\\
main\_id& The main identifier of the host star, as provided by SIMBAD/K2/EPIC/Gaia catalogs.& STRING\\
discovery\_method& The discovery method of the planet.& STRING\\
binary& String labeling the binary host star, if any.& STRING\\
ra\_off& The J2000 right ascension in degrees, as provided by SIMBAD/K2/EPIC/Gaia catalogs.& FLOAT\\
dec\_off& The J2000 declination in degrees, as provided by SIMBAD/K2/EPIC/Gaia catalogs.& FLOAT\\
Status& The string AXDXEXCX showing the status of the planet in all source catalogs.& STRING\\
Status\_string& Most probable status of the planet.& STRING \\
confirmed& Number of 3 values in the Status column.& INTEGER\\
yod& Year of the discovery of the planet.& INTEGER\\
alias& Known aliases for the host star.& STRING \\
catalog& List of catalogs in which the target appears.& STRING\\
MismatchFlagHost& Flag displaying the probable binary duplicate. &INTEGER\\

  \hline  

\end{longtable}

\section{VO metadata and resource registration details}
\label{app:vodetails}

From the descriptions in Table~\ref{tab:catdesc} and the physical mapping of the \textit{Type} column from that table into the actual database a set of metadata information has been retrieved. 
The \textit{Meaning} column content was directly entered in the TAP \textit{TAP\_SCHEMA} column description field, the same description was also used to identify the possible \textit{units} for the column values and the UCD \citep{2018ivoa.spec.0527M} vocabulary terms to annotate them. A summary of these metadata can be seen in Table~\ref{tab:tapmeta}, where the \textit{units} are based on the VOUnits \citep{2014ivoa.spec.0523D} standard and the \textit{UCD}s are build out of suggestions from the CDS UCD suggest service\footnote{http://cds.u-strasbg.fr/UCD/cgi-bin/descr2ucd} consumed through the TASMAN application developed at the IA2 \footnote{http://ia2.inaf.it} data center. 

\begin{longtable}{l|p{3cm}|l}
\caption{Derived metadata information for the columns as exposed
through the TAP service.\label{tab:tapmeta}}
\\
  \hline\hline
  column\_name & unit & ucd\\\hline 
name & & meta.id\\
host & & \\
letter & & \\
ra\_off & deg & pos.eq.ra;meta.main\\
dec\_off & deg & pos.eq.dec;meta.main\\
mass & MJup & phys.mass\\
mass\_max & MJup & stat.error;phys.mass;stat.max\\
mass\_min & MJup & stat.error;phys.mass;stat.min\\
mass\_url & & meta.bib.bibcode\\
msini & MJup & phys.mass;stat.min\\
msini\_max & MJup & phys.mass;stat.min;stat.max\\
msini\_min & MJup & phys.mass;stat.min;stat.min\\
msini\_url & & meta.bib.bibcode\\
bestmass & MJup & phys.mass\\
bestmass\_min & MJup & stat.error;phys.mass;stat.min\\
bestmass\_max & MJup & stat.error;phys.mass;stat.max\\
bestmass\_url & & meta.bib.bibcode\\
mass\_prov & & meta.code;phys.mass\\
p & d & time.period\\
p\_max & d & stat.error;time.period;stat.max\\
p\_min & d & stat.error;time.period;stat.min\\
p\_url & & meta.bib.bibcode\\
r & RJup & phys.size.radius\\
r\_max & RJup & stat.error;phys.size.radius;stat.max\\
r\_min & RJup & stat.error;phys.size.radius;stat.min\\
r\_url & & meta.bib.bibcode\\
a & au & phys.size.smajAxis\\
a\_max & au & phys.size.smajAxis;stat.max\\
a\_min & au & phys.size.smajAxis;stat.min\\
a\_url & & meta.bib.bibcode\\
e & & src.orbital.eccentricity\\
e\_max & & stat.error;src.orbital.eccentricity;stat.max\\
e\_min & & stat.error;src.orbital.eccentricity;stat.min\\
e\_url & & meta.bib.bibcode\\
i & deg & src.orbital.inclination\\
i\_max & deg & stat.error;src.orbital.inclination;stat.max\\
i\_min & deg & stat.error;phys.mass;stat.min\\
i\_url & & meta.bib.bibcode\\
main\_id & & meta.id;meta.main\\
binary & & meta.code.class\\
discovery\_method & & \\
status & & meta.code\\
status\_string & & \\
confirmed & & meta.code\\
yod & yr & time.epoch\\
alias & & meta.id\\
catalog & & \\
update\_time & & \\
mismatch\_flag\_host & & meta.code\\
  \hline
\end{longtable}

With the above content the catalog has been registered as a VO resource having IVOID \texttt{ivo://ia2.inaf.it/catalogues/exomercat} and served by the TAP service, itself registered in the VO with IVOID  \texttt{ivo://ia2.inaf.it/tap/projects}. The former identifier should persistently represent the Exo-MerCat catalog and its evolution in time, the latter the TAP service that currently deploys Exo-MerCat content.






\biboptions{authoryear}
\bibliographystyle{elsarticle-harv}
 \bibliography{exobib}





\end{document}